\documentclass[trackchanges, twocolumn]{aastex7} 
\usepackage{amsmath}	
\usepackage{amssymb}	
\usepackage{longtable}
\usepackage{booktabs}
\usepackage{scalerel}
\usepackage{hyperref}

\mathchardef\mhyphen="2D

\newcommand{\oii}{O\,{\sc ii}}
\newcommand{\oiii}{O\,{\sc iii}}

\mathchardef\mhyphen="2D

\def\ha{H$\alpha$}
\def\hb{H$\beta$}

\def\hii{H\,{\sc ii}}

\def\nii{N\,{\sc ii}}

\def\oiii{O\,{\sc iii}}

\def\Q0059{Q0059--2735}

\def\S2S3{S2S3}

\definecolor{blk}{rgb}{0.0,0.0,0.0}
\definecolor{red}{rgb}{0.75,0.0,0.0}
\definecolor{yel}{rgb}{0.65,0.65,0.0}
\definecolor{grn}{rgb}{0.0,0.75,0.0}
\definecolor{blu}{rgb}{0.0,0.0,0.75}
\definecolor{gry}{rgb}{0.75,0.75,0.75}

\def\nh{\ifmmode n_\mathrm{\scriptscriptstyle H} \else $n_\mathrm{\scriptscriptstyle H}$\fi}
\def\ne{\ifmmode n_\mathrm{\scriptstyle e} \else $n_\mathrm{\scriptstyle e}$\fi}
\def\Te{\ifmmode T_\mathrm{\scriptstyle e} \else $T_\mathrm{\scriptstyle e}$\fi}
\def\Qh{\ifmmode Q_\mathrm{\scriptstyle H} \else $Q_\mathrm{\scriptstyle H}$\fi}
\def\Uh{\ifmmode U_\mathrm{\scriptstyle H} \else $U_\mathrm{\scriptstyle H}$\fi}
\def\Nh{\ifmmode N_\mathrm{\scriptstyle H} \else $N_\mathrm{\scriptstyle H}$\fi}
\def\Nhi{\ifmmode N_\mathrm{\scriptstyle HI} \else $N_\mathrm{\scriptstyle HI}$\fi}
\def\Uhhp{\ifmmode U_\mathrm{\scriptstyle H,HP} \else $U_\mathrm{\scriptstyle H,HP}$\fi}
\def\Nhhp{\ifmmode N_\mathrm{\scriptstyle H,HP} \else $N_\mathrm{\scriptstyle H,HP}$\fi}
\def\Uhvhp{\ifmmode U_\mathrm{\scriptstyle H,VHP} \else $U_\mathrm{\scriptstyle H,VHP}$\fi}
\def\Nhvhp{\ifmmode N_\mathrm{\scriptstyle H,VHP} \else $N_\mathrm{\scriptstyle H,VHP}$\fi}
\def\Nion{\ifmmode N_\mathrm{\scriptstyle ion} \else $N_\mathrm{\scriptstyle ion}$\fi}

\def\Zsun{\ifmmode {\rm Z}_{\odot} \else $Z_{\odot}$\fi}
\def\Msun{\ifmmode {\rm M}_{\odot} \else M$_{\odot}$\fi}
\def\kms{\ifmmode {\rm km~s}^{-1} \else km~s$^{-1}$\fi}
\def\Lya{\ifmmode {\rm Ly}\alpha \else Ly$\alpha$\fi}
\def\Lyb{\ifmmode {\rm Ly}\beta \else Ly$\beta$\fi}
\def\Lyg{\ifmmode {\rm Ly}\gamma \else Ly$\gamma$\fi}
\def\Lyd{\ifmmode {\rm Ly}\delta \else Ly$\delta$\fi}
\def\neaod{\ifmmode n_\mathrm{\scriptscriptstyle AOD} \else $n_\mathrm{\scriptscriptstyle AOD}$\fi}
\def\necrit{\ifmmode n_\mathrm{\scriptstyle cr} \else $n_\mathrm{\scriptstyle cr}$\fi}
\def\ncr{\ifmmode n_\mathrm{\scriptstyle cr} \else $n_\mathrm{\scriptstyle cr}$\fi}
\def\nepi{\ifmmode n_\mathrm{\scriptscriptstyle PI} \else $n_\mathrm{\scriptscriptstyle PI}$\fi}
\def\gtorder{\mathrel{\raise.3ex\hbox{$>$}\mkern-14mu\lower0.6ex\hbox{$\sim$}}}
\def\ltorder{\mathrel{\raise.3ex\hbox{$<$}\mkern-14mu\lower0.6ex\hbox{$\sim$}}}

\def\vro{\ifmmode v_\mathrm{\scriptscriptstyle 1, \scriptstyle r} \else $v_\mathrm{\scriptscriptstyle 1, \scriptstyle r}$\fi}
\def\vrc{\ifmmode v_\mathrm{\scriptscriptstyle 2, \scriptstyle r} \else $v_\mathrm{\scriptscriptstyle 2, \scriptstyle r}$\fi}
\def\vzo{\ifmmode v_\mathrm{\scriptscriptstyle 1, \scriptstyle z} \else $v_\mathrm{\scriptscriptstyle 1, \scriptstyle z}$\fi}
\def\vzc{\ifmmode v_\mathrm{\scriptscriptstyle 2, \scriptstyle z} \else $v_\mathrm{\scriptscriptstyle 2, \scriptstyle z}$\fi}

\newcommand{\FWHMB}{FWHM$_B$}
\newcommand{\FWHMN}{FWHM$_N$}

\newcommand{\Vmax}{$V_{\textrm{max}}$}

\newcommand{\Mstar}{\text{M}$_{\star}$}

\newcommand{\MdotOut}{$\dot{\text{M}}_\text{out}$}

\newcommand{\SigmaSFR}{$\Sigma_\mathrm{SFR}$}

\def\ZStar{\ifmmode {\rm Z}_\text{stars} \else $Z_\text{stars}$\fi}
\def\ZGas{\ifmmode {\rm Z}_\text{gas} \else $Z_\text{gas}$\fi}

\revised{\today}

\begin{document}

\title{Scaling Relations of Galactic Outflows Across Cosmic Time: New Insights from Cosmic Noon}

\author[]{Tiffany Liou}
\email{tiffanyliou2025@u.northwestern.edu}
\affiliation{Department of Physics and Astronomy, Northwestern University, 2145 Sheridan Road, Evanston, IL, 60208, USA}
\affiliation{Center for Interdisciplinary Exploration and Research in
Astrophysics (CIERA), 1800 Sherman Avenue,
Evanston, IL, 60201, USA}

\email{tiffanyliou2025@u.northwestern.edu}

\author{Xinfeng Xu}
\email{<xinfeng.xu@email.address>}
\affiliation{Department of Physics and Astronomy, Northwestern University,
2145 Sheridan Road, Evanston, IL, 60208, USA}
\affiliation{Center for Interdisciplinary Exploration and Research in
Astrophysics (CIERA), 1800 Sherman Avenue,
Evanston, IL, 60201, USA}

\author{Allison L. Strom}
\email{<allison.strom@northwestern.edu>}
\affiliation{Department of Physics and Astronomy, Northwestern University,
2145 Sheridan Road, Evanston, IL, 60208, USA}
\affiliation{Center for Interdisciplinary Exploration and Research in
Astrophysics (CIERA), 1800 Sherman Avenue,
Evanston, IL, 60201, USA}

\author{Nathalie A. Korhonen Cuestas}
\email{<NathalieKorhonenCuestas2029@u.northwestern.edu>}
\affiliation{Department of Physics and Astronomy, Northwestern University,
2145 Sheridan Road, Evanston, IL, 60208, USA}
\affiliation{Center for Interdisciplinary Exploration and Research in
Astrophysics (CIERA), 1800 Sherman Avenue,
Evanston, IL, 60201, USA}

\author{Claude-André Faucher-Giguère}
\email{<cgiguere@northwestern.edu>}
\affiliation{Department of Physics and Astronomy, Northwestern University,
2145 Sheridan Road, Evanston, IL, 60208, USA}
\affiliation{Center for Interdisciplinary Exploration and Research in
Astrophysics (CIERA), 1800 Sherman Avenue,
Evanston, IL, 60201, USA}

\author{Tim B. Miller}
\email{<timothy.miller@northwestern.edu>}
\affiliation{Center for Interdisciplinary Exploration and Research in
Astrophysics (CIERA), 1800 Sherman Avenue,
Evanston, IL, 60201, USA}

\author{Gwen Rudie}
\email{<gwen@carnegiescience.edu>}
\affiliation{The Observatories of the Carnegie Institution for Sciences, 813 Santa Barbara Street, Pasadena, CA 91101, USA}

\author{Ryan F. Trainor}
\email{ryan.trainor@fandm.edu}
\affiliation{Department of Physics and Astronomy, Franklin \& Marshall College, 637 College Avenue, Lancaster, PA 17603, USA}

\author{Naveen A. Reddy}
\email{naveenr@ucr.edu}
\affiliation{Department of Physics and Astronomy, University of California, Riverside, 900 University Avenue, Riverside, CA 92521, USA}

\author{Charles C. Steidel}
\email{ccs@astro.caltech.edu}
\affiliation{Cahill Center for Astronomy and Astrophysics, California Institute of Technology, MC 249-17, Pasadena, CA 91125, USA}

\author{Yuguang Chen}
\email{yuguangchen@cuhk.edu.hk}
\affiliation{Department of Physics, The Chinese University of Hong Kong, Shatin, N.T., Hong Kong SAR, People’s Republic of China}

\begin{abstract}
Galactic outflows play a significant role in regulating galaxy evolution. Scaling relations between stellar mass (\Mstar), star formation (SFR), and outflow properties have been extensively studied at low redshifts ($z<1$) but less so beyond $z \sim 2$. We construct a joint sample of 387 galaxies from $z \sim$ 0--9, including 98 new Cosmic Noon galaxies from the Keck Baryonic Structure Survey and the Keck Lyman Continuum Spectroscopic Survey. Using high signal-to-noise emission lines (SNR $>$ 50 for \ha\ or [\oiii] $\lambda5007$) from Keck/MOSFIRE spectra, we detect warm-ionized outflows by decomposing lines into narrow and broad components. With the joint sample, we explore the redshift evolution of outflow scaling relations. On average, outflows at Cosmic Noon have higher maximum velocities than those at low$-z$ by up to a factor of 3 for a fixed \Mstar, SFR, or SFR surface density. They also have higher mass outflow rates for a fixed \Mstar. Despite faster and stronger outflows, there is no evolution in the mass loading factor for a fixed \Mstar. We find evidence for galactic fountains, as the majority of outflowing gas is recycled at a radius of $\sim$0.03 R$_\textrm{vir}$. We also constrain the overall outflow occurrence rate in our galaxy sample to be at least 30$\%$ when taking galaxy orientation and outflow geometry into account. By analyzing the largest sample of warm-ionized outflows at Cosmic Noon to date and compiling large galaxy samples across all redshifts, we present a comprehensive analysis of galactic outflows throughout cosmic time.

\end{abstract}


\section{Introduction} 
\label{sec:intro}
The mass growth and chemical evolution of galaxies are directly influenced by feedback mechanisms such as galactic-scale outflows. Outflows, driven by active galactic nuclei (AGNs), supernova explosions (SNe), radiative pressure, stellar winds, and intense star formation, inject energy and momentum into the interstellar medium (ISM). In turn, these outflows regulate the mass growth and star-formation rate (SFR) of galaxies by expelling gas into the circumgalactic medium (CGM) or delaying the accretion of cold gas onto the central galaxy (e.g., \citealt{Thompson24} and references therein). To understand the full picture of galaxy evolution, it is essential to link the effects of large-scale outflows to galaxy properties.

Galactic outflows are multi-phase and are observed at multiple wavelengths \citep[e.g., see][for reviews]{Rupke18,  Veilleux20, Thompson24}. The hot phase ($T \sim$ 10$^7$ K) that results from starbursts or SNe is observed in the X-ray. Ultraviolet and optical observations trace the warm ($T \sim$ 10$^4$ K) and cool ($T \sim$ 10$^3$ K) phases from photoionized gas in star-forming regions. Additionally, radio emission traces cooler to cold gas ($T \sim$ 10$^2$ K). Understanding the role of each outflow phase is important for investigating feedback.

Cosmic Noon ($z\sim 2-3$) marks the peak of star formation and black hole activity in the Universe \citep{Madau14}. Galaxies at Cosmic Noon thus provide a unique opportunity to understand feedback effects when many galaxies were growing rapidly \citep{Davies19, Forster19, Madau14}. One method of detecting outflowing gas in the warm-ionized phase is to use broadened emission lines, which arise from the larger velocity dispersion of this warm gas relative to the ISM. The tracers of both star formation and outflows in the warm-ionized phase are rest-optical emission lines such as \hb, [\oiii] $\lambda\lambda$4959, 5007, \ha, and [\nii] $\lambda\lambda$6548, 6583. 

Quantifying the scaling relations between starburst-driven outflows and galaxy properties has long been an area of active research. Scaling relations have been studied extensively in the local Universe, and tight correlations between outflow and galaxy properties have been established in the local ($z<1$) Universe \citep[e.g., see][]{Heckman15, Xu22a, Marasco23, Schroetter24, Peng2025, Xu25}. However, medium $z \sim 2-3$) to high ($z \gtrsim 4$) redshift studies using individual galaxies show mixed trends and have not converged on a consistent picture. For example, \cite{Weldon24} concluded that there are no correlations between the maximum outflow velocity (\Vmax) and any galaxy property using a sample size of 33 galaxies; however, \cite{Llerena23} found a weak positive correlation between \Vmax\ and the star formation surface density (\SigmaSFR) for 23 galaxies. At $z\sim3-9$, \cite{XuY25} finds a weak correlation between \Vmax\ and stellar mass (\Mstar) for a sample of 30 galaxies, but not with other galaxy properties. Interestingly, \cite{Weldon24} and \cite{Davies19} find clear correlations between outflow properties when using stacked spectra to increase the signal-to-noise ratio (SNR) of diagnostic emission lines. 

The mixed conclusion at higher redshifts stems from limited sample sizes for the following reasons. First, because winds are quantified by their velocity dispersion relative to the surrounding ISM, higher spectral resolution ($R>2000$) is required to robustly separate their emission line signatures \citep[e.g.,][]{XuY25}. Second, emission lines must have high SNR in order to detect outflows at high significance; e.g., \cite{Xu25} concluded that SNR $>$ 50 in diagnostic emission lines is needed for 3$\sigma$ outflow detections. However, obtaining spectra with both high spectral resolution and high SNR is observationally expensive, especially for distant and/or faint galaxies. One method commonly used in high-redshift outflow studies is spectral stacking. While this approach generally achieves higher SNR, it has the caveat that it blends the kinematic structures of individual galaxies, potentially introducing uncertainties, such as artificially extending broad emission components. Overall, it is crucial to study outflows in large samples of individual medium- to high-redshift galaxies to fully understand the dependence of outflows on galaxy properties. 

To study outflows from individual galaxies, long exposure times are necessary to obtain deep spectra with high SNR. Because diagnostic emission lines such as [\oiii] and \ha\ are redshifted to the infrared at Cosmic Noon, a powerful infrared instrument with sufficient spectral resolution ($R>2000$) on a large telescope is needed. With these challenges in mind, we conduct such an analysis using Keck/MOSFIRE \citep{McLean12}. We build a large sample of galactic outflows, on top of previous literature, using the high--SNR spectra from the Keck Baryonic Structure Survey \citep[KBSS;][]{Steidel14, Strom17} and the Keck Lyman Continuum Spectroscopic Survey \citep[KLCS;][]{Steidel18}. We explore the scaling relations between warm-ionized outflows and galaxy properties, as well as any redshift evolution of these trends. A brief overview of the paper is as follows. In Section \ref{sec:data}, we describe the KBSS and KLCS sample selection. Then, we detail our outflow selection and outflow measurements in Section \ref{sec:methods}. Our results for outflow and galaxy scaling relations are presented in Section \ref{sec:results}, and the implications of our findings are discussed in Section \ref{sec:discuss}. Throughout the paper, we assume a cosmological model with $H_0$ = 70 km/s/Mpc, $\Omega_\Lambda$ = 0.7, and $\Omega_m$ = 0.3.

\section{Data and Measurements} 
\label{sec:data}

\subsection{Observations} 
\label{sec:observations}
KBSS targets star-forming galaxies at redshifts $z \sim 1.5-3.5$ in 15 quasar fields over a total area of $\sim$ 0.24 deg$^2$. The survey combines rest-UV and rest-optical imaging and spectroscopy. The spectroscopy is obtained with Keck/LRIS \citep{Oke95} and Keck/MOSFIRE \citep{McLean12}, which provide a medium spectral resolution of R $\sim$ 3300, 3690, and 3620 in the $J$, $H$, and $K$ bands, respectively \citep{Steidel14}. For this study, we focus on MOSFIRE spectra. By design, the majority of strong rest-optical emission lines \hb, $[$\oiii$]$ $\lambda\lambda$4959, 5007, \ha, and $[$\nii$]$ $\lambda\lambda$6548, 6583 are contained in the $J$, $H$, and $K$ bands. Most target galaxies are selected based on their rest-UV colors, whereas some are selected using the $R-K$ color to avoid a bias against more dusty star-forming galaxies \citep{Strom17}. In this study, we include 1376 galaxies from KBSS.

One advantage of the KBSS survey design is its observation depth and high--SNR spectra. Observations of target galaxies are, on average, repeated on multiple masks. The average exposure time of each galaxy ranges from 1.5 to 7.5 hours per band, depending on the number of observations. For comparison, another benchmark survey of galaxies at Cosmic Noon, the MOSFIRE Deep Evolution Field (MOSDEF) survey \citep{Kriek15, Freeman19}, has exposure times of 1 to 3 hours per band with the same instrument.On average, each galaxy has an exposure time of 5.22 hours.

In addition to the 15 KBSS quasar fields, we include 274 galaxies from 6 KLCS fields. KLCS targets Lyman-break star-forming galaxies at $z = 2.8-3.5$. KLCS also obtains imaging and spectroscopy using LRIS and MOSFIRE. For the purpose of this study, we include only spectra observed using MOSFIRE across the $J$, $H$, and $K$ bands. On average, each object has a total exposure time of 7.66 hours. We refer the reader to \cite{Steidel14}, \cite{Strom17}, and \citep{Steidel18} for detailed descriptions of data reduction and extraction of KBSS and KLCS.

In total, there are 1650 galaxies in the parent KBSS+KLCS sample. Then, we select a subset of high--SNR galaxies (N=323) where SNR(\ha)$>$50 or SNR([\oiii])$>$50 in Section \ref{sec:High-SNR Sample}. Finally, we identify the outflow sample (N=98) in Sections \ref{sec: gaussian} and \ref{sec: stat tests}.

\subsection{SED fitting}
\label{sec:SED}

 We performed spectral energy distribution (SED) fits for each galaxy using the Bayesian Analysis of Galaxies for Physical Inference and Parameter EStimation (\texttt{BAGPIPES}) package \citep{Carnall18} and available broadband photometry. All galaxies in the final outflow sample (N=98) have SED fits performed (see Section \ref{sec: occurence}). \texttt{BAGPIPES} uses the nested sampling algorithm from \texttt{Nautilus} to sample prior distributions and infer physical quantities \citep{Skilling2006, nautilus}. Nebular emission lines are not included in the SED model because the flux contribution from nebular emission lines has been subtracted from all photometric measurements, with the exception of the Westphal field, using measurements from the Keck/MOSFIRE spectra. Contributions from the nebular continuum are, however, considered.
 
 In total, KBSS and KLCS fields have photometry in up to 14 and 19 bands, respectively. The number of photometric data points varies across fields, but all 15 KBSS and 6 KLCS fields have photometry in at least 8 bands. We refer the reader to \cite{Strom17}, \cite{Theios2019}, and \cite{KorhonenCuestas2025} for more details on the available photometry for KBSS and \cite{Pahl2023} for KLCS.

For a more detailed description of the SED fitting procedure, we again refer the reader to \cite{KorhonenCuestas2025}; here, we provide a brief summary. For each galaxy, we vary stellar metallicity ($Z_\star$), ionization parameter ($U$), and dust attenuation ($A_V$). For each respective quantity, we provide a uniform prior of $Z_\star$ $\in$ [0.05, 0.6]$Z_\odot$ (see e.g., \cite{Rogers2025}), log($U$) $\in$ [--3.5,--2.5], and $A_V$ $\in$ [0,2]. We allow log($U$) to vary between --3.5 and --2.5. This is consistent with the expected range of ionization parameters for KBSS galaxies. For example, \cite{Strom18} infers that log($U$) for $z\approx2-3$ KBSS galaxies lies between --2.93 and --2.58; others using different photometric models suggests that log($U$) ranges between [--3.1, --2.5] \citep[see e.g.,][]{steidel16, Strom17}.

In the model fits, we assume the SMC \citep{Gordon2003} dust reddening curve and adopt a non-parametric, continuity star formation history \citep[SFH;][]{Leja2019}. We fit for the change in SFR across each time step, with increments beginning from 0, 10, 100, 250, 500, 1000, 1500, 2000, 2500, 3000, and 3500 Myr prior to observation with a Student's t prior on the change in SFR, which generally produces a smoother SFH. It is important to note that the default initial mass function (IMF) in \texttt{BAGPIPES} is Kroupa-like, but we rescale the resulting stellar mass fits by a factor of +0.924 \citep{Madau14} to obtain the \cite{Chabrier03} IMF to be consistent with other analyses \citep[e.g.,][]{Freeman19, Llerena23, Carniani24, XuY25}. In the remainder of the analysis, the SED parameters we use are the model stellar continua and the stellar mass (\Mstar).

\subsection{High--SNR Galaxy Sample Selection}
\label{sec:High-SNR Sample}
To detect the outflow signatures (Section \ref{sec: gaussian}), we focus on the strongest nebular emission lines available at $z \sim 2-3$ (e.g., \ha, \hb, [\oiii], and [\nii]). The total KBSS+KLCS sample consists of 1650 galaxies, where 841 galaxies have \ha\ detections and 1000 galaxies have [\oiii] $\lambda$5007 detections. Following \cite{Xu25}, we filter all KBSS+KLCS galaxies based on whether the \ha\ or [\oiii] $\lambda$5007 emission lines have SNR $>$ 50. After applying the SNR criteria, our high--SNR galaxy sample has 323 galaxies, of which 44 are from KLCS and 279 are from KBSS. All 44 KLCS galaxies have [\oiii] coverage, while 2 of 44 have \ha\ due to redshift coverage ($z~3-3.4$). For KBSS, 267 galaxies have [\oiii] coverage and 257 galaxies have \ha. For the high--SNR sample, 138/259 galaxies have SNR(\ha) $>$ 50 and 263/311 galaxies have SNR([\oiii]) $>$ 50. There are 78 galaxies that satisfy both SNR criteria.

\section{Methods and Data Analysis}
\label{sec:methods}

\subsection{Outflow Extraction through Gaussian Fitting}
\label{sec: gaussian}
To select the outflow sample (i.e., galaxies with detectable outflows), we follow the procedure of previous works to perform a double-Gaussian fit composed of a broad component that traces the kinematics of the outflowing gas and a narrow component that traces the gas in the ISM \citep[e.g.,][]{Freeman19, Weldon24}. All of our lines of interest (\hb, [\oiii] $\lambda\lambda 4959,5007$, \ha, [\nii] $\lambda\lambda6548, 6583$) are fit simultaneously. For the remainder of this paper, "[\oiii]" is the shorthand notation for [\oiii] $\lambda5007$. 

The fit parameters for the narrow and broad components are motivated by physical constraints and assumptions made by previous studies \citep[e.g.,][]{Genzel11, Weldon24, Newman12a}. We assume that all fitted lines for each galaxy share the same ISM velocity shift ($\Delta v_N$) compared to the systematic velocity of the galaxy, broad component velocity shift ($\Delta v_B$), and full width half maxima (FWHM) of the narrow and broad components (\FWHMN\ and \FWHMB). For our sample, \ha\ and [\oiii] lie in different spectroscopic bands, so the spectral resolution is slightly different. We tested the effects of the resolution difference by allowing the two lines to have slightly different line centers and FWHM. This resulted in only minor changes in our final solutions. Thus, we proceed with the method where $\Delta v_N$, $\Delta v_B$, \FWHMN, and \FWHMB\ are the same for all line transitions within each galaxy. 

\begin{figure*}
    \centering
    \includegraphics[width=0.5\linewidth]{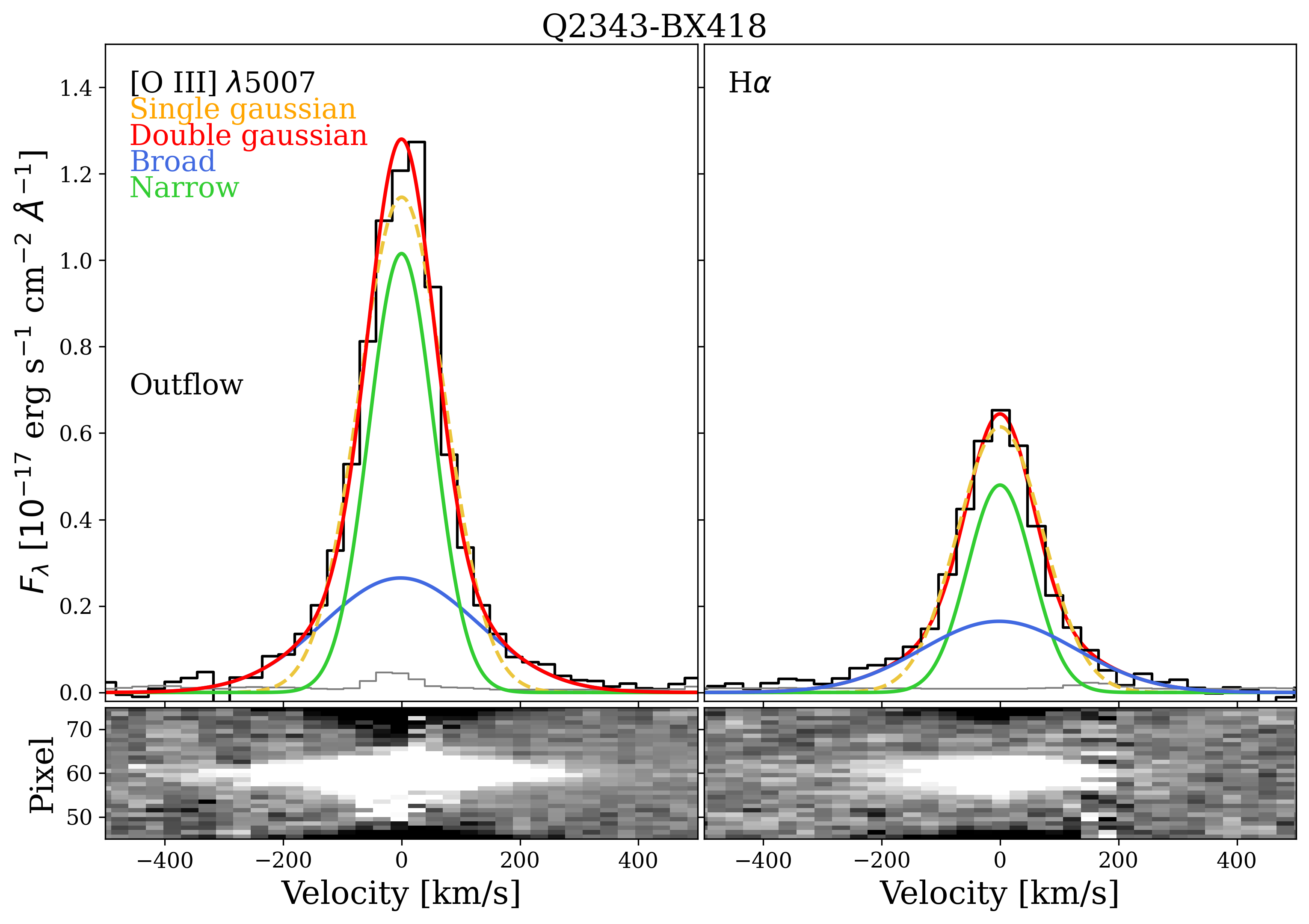}%
    \includegraphics[width=0.5\linewidth]{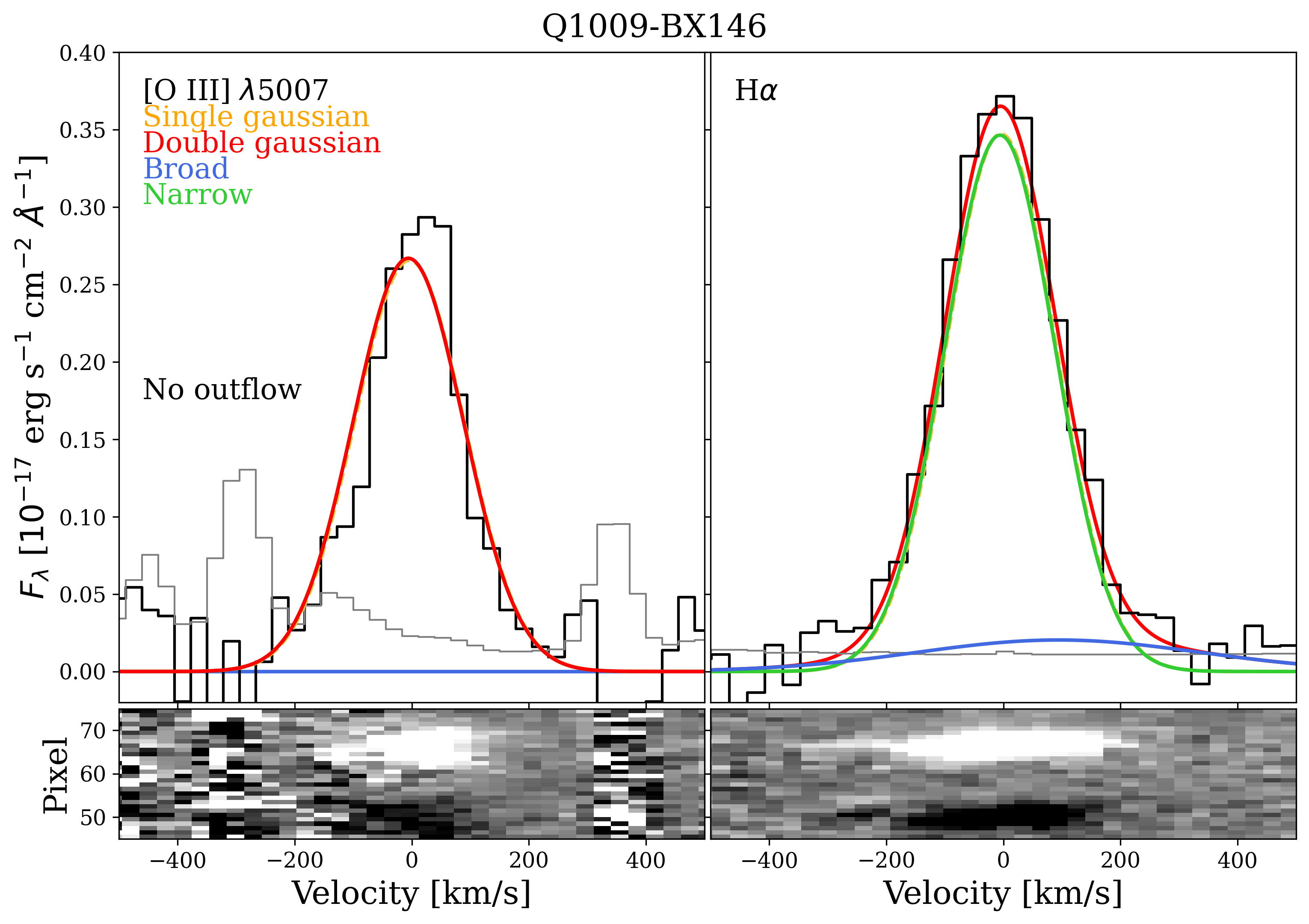}

    \caption{Sample emission-line fits for Q2343-BX418 (left) and Q1009-BX146 (right). We show the best-fit narrow (green), broad (blue), composite double-Gaussian (red), and single-Gaussian (orange) fits to the [\oiii] $\lambda$5007 and \ha\ emission lines. The noise on the spectra is represented by the light grey lines. The 2D spectra are shown below. Outflow components in both lines are detected in Q2343-BX418, as a double Gaussian can better describe the line profiles than a single Gaussian. No outflows are detected for Q1009-BX146 because the single- and double-Gaussian fits are comparable. 
    }
    \label{fig:2G fit}
\end{figure*}

Additionally, we tie the flux ratios of certain doublet lines, with [\nii] $\lambda$6583/[\nii] $\lambda$6548 = 2.93 and [\oiii] $\lambda$5007/[\oiii] $\lambda$4959 = 2.98 \citep{Osterbrock06}. Overall, for all six lines considered in the double-Gaussian fit, there are a total of 12 free parameters: $\Delta v_N$, $\Delta v_{B}$, \FWHMB, \FWHMN, 4 independent broad line flux amplitudes A$_B$, and 4 independent narrow line flux amplitudes A$_N$. The single-Gaussian fit has 6 free parameters: $\Delta v_N$, \FWHMN, and 4 independent line flux amplitudes. Each Gaussian profile is defined as follows: 
\begin{equation}
    G(v, \mathrm{A}_i, \Delta v_i, \sigma_i) =
    \frac{\mathrm{A}_{i,j}}{\sqrt{2\pi\sigma_i}}
    \exp\!\left( -\frac{1}{2}\frac{ \left( v - \Delta v_i \right)^2 }{ \sigma_i^2 } \right)
\end{equation}
where the subscript $i$ is N or B depending on the component being fit, the subscript $j$ is the emission line being fit, $v$ is the velocity, and $\sigma_i$ is the velocity dispersion of the component, given by $\sigma_i = \textrm{FWHM}_i/2\sqrt{2\textrm{ln2}}$. 

For the double-Gaussian fit, we apply different constraints to the broad and narrow components (Table \ref{tab:fitting_params}). The main distinction between the components is their FWHM, with the broad components having a larger width because the outflowing gas has a larger physical velocity dispersion. As narrow and broad components are difficult to disentangle, it is a common approach to set boundaries on \FWHMN\ and \FWHMB\ such that the two widths do not overlap. For example, \cite{Freeman19} limits \FWHMN\ to be less than 275 \kms\ based on the velocity distributions of their sample. For our sample, we set \FWHMN\ $<$ 250 \kms\ based on the ISM velocity dispersion estimated using a size-mass relation \citep[see Figure 2 in][]{Bezanson2011}. For reference, galaxies in the outflow sample have stellar masses between the range $10^{8}<$ \Msun $< 10^{11}$. The majority of our galaxies have a velocity dispersion $\sigma_{\rm{N}}$ $<$ 100 \kms\ (corresponding to \FWHMN\ $<$ 235 \kms), suggesting that a higher upper limit on \FWHMN\ is not necessary. It is important to note that our choice of FWHM boundary at 250 \kms\ systematically excludes outflows at lower velocities.

We also require that the narrow component contributes at least 20$\%$ of the total emission line flux amplitude and has an upper limit of 100$\%$ \citep{Freeman19}. Accordingly, the broad component flux amplitude is limited to contribute up to 80$\%$ of the total flux. We restrict $\Delta v_N$ to be within $\pm$ 30 \kms, approximately $\pm$ 1 pixel, of the galaxy's systemic velocity, measured using spectroscopic redshifts reported in \cite{steidel16}. As the outflowing gas is typically blue-shifted from the systemic velocity of the galaxy, we allow $\Delta v_B$ to shift in a greater range ($-100 < \Delta v_B < +100$) \kms. 

Before fitting the lines, we first subtract the best-fit SED continuum (Section \ref{sec:SED}) from the observed spectrum. The SED continuum was normalized and rescaled by the ratio of the average observed flux to the continuum in a wavelength region, excluding any emission lines. Then we identify a fitting region of $\pm$500 \kms\ surrounding each emission line center to avoid overfitting and noise from OH sky lines. This fitting window also minimizes the blending of \ha\ features with those of [\nii]. The fit parameters were derived using the bounded limited-memory Broyden-Fletcher-Goldfarb-Shanno (L-BFGS-B) method in \texttt{minimize} from the \texttt{scipy.optimize} \citep{2020SciPy-NMeth} subpackage in Python, which uses a gradient descent optimization method for non-linear problems. The corresponding uncertainties for each parameter were obtained using the output covariance matrix.  

\begin{table}[ht]
\centering
\caption{Gaussian Fitting Parameters for Narrow vs. Broad Components}\label{tab:fitting_params}
\begin{tabular}{l @{\hspace{1cm}} c}
\toprule
\textbf{Parameter} & \textbf{Constraint} \\
\midrule
FWHM$_N$       & 0--250 km\,s$^{-1}$ \\
$A_N$          & 20--100\% of total line flux\\
$\Delta v_N$   & $-30$ to +$30$ km\,s$^{-1}$ \\
\hline
FWHM$_B$       & 250--1000 km\,s$^{-1}$ \\
$A_B$          & 0--80\% of total line flux\\
$\Delta v_B$   & $-100$ to +$100$ km\,s$^{-1}$ \\
\bottomrule
\end{tabular}
\end{table}

\subsection{Statistical Verification of Outflows}
\label{sec: stat tests}
For each galaxy, we compare the double-Gaussian results to the single-Gaussian fits using statistical tests to ensure our outflow detection is significant. If a galaxy has outflow signatures, its emission lines should be best described by a combination of narrow and broad Gaussian components. Thus, we implement three statistical tests.

First, we use an $F$-test to determine whether the statistical variance of the double-Gaussian fit is less than the single-Gaussian fit (i.e., if the emission line is entirely reflective of the ISM gas):
\begin{equation}
\label{eq:F-test}
    F = \frac{(\chi_1^2 -\chi_2^2)/(p_1-p_2)}{\chi_2^2/(n-p_2)},
\end{equation}
where $\chi_1$ and $\chi_2$ are the chi-square values for the single and double Gaussian fits, respectively. The statistic is also weighted by the number of fitting parameters, $p_1$ and $p_2$, and by the number of points used in the fit, $n$. For the single and double Gaussian fits, $p_1=6$ and $p_2=12$. The $F$-score is compared to a critical value set by the theoretical $F$-distribution based on the significance parameter $\alpha=0.05$ and the degrees of freedom allowed in the fit. If the $F$-score is greater than the critical value, we reject the null hypothesis, meaning that a double-Gaussian model fit is needed to describe the observed emission line.

Second, we use the Bayesian information criterion (BIC) as an additional filter: \begin{equation}
    \label{eq:BIC}
    \mathrm{BIC} = \chi^2 + k\ln(n)
\end{equation}
where $\chi^2$ is the chi-squared of the fit, $k$ is the number of fitted parameters, and $n$ is the number of points considered. The BIC scores for the single- and double-Gaussian models are then compared. We implement a similar criterion as previous studies \citep[e.g.,][]{Swinbank19, Weldon24} so that $\Delta$BIC = BIC$_{\textrm{single}}$ -- BIC$_{\textrm{double}}$ $\gtrsim$ 10 is an indication that a double-Gaussian fit is more significant than a single-Gaussian one. 

Finally, to ensure our broad outflow components are physical and not an artifact of the fitting procedure, we require $A_B > 3 A_{B,\textrm{err}}$, i.e., the broad component's amplitude is 3 times greater than its uncertainty. Examples of our double-Gaussian fits are shown in Figure \ref{fig:2G fit}. Our final outflow sample after these statistical tests consists of 98 galaxies at $z \sim 1.5 - 3.3$. Further analysis of the outflow sample and comparisons to the high--SNR sample are described later in Section \ref{sec: occurence}. We present the Gaussian fit results for each galaxy in the Appendix. (Please find the table in the ApJ version of the paper.)

\begin{figure}[h]
    \centering
    \includegraphics[width=1 \linewidth]{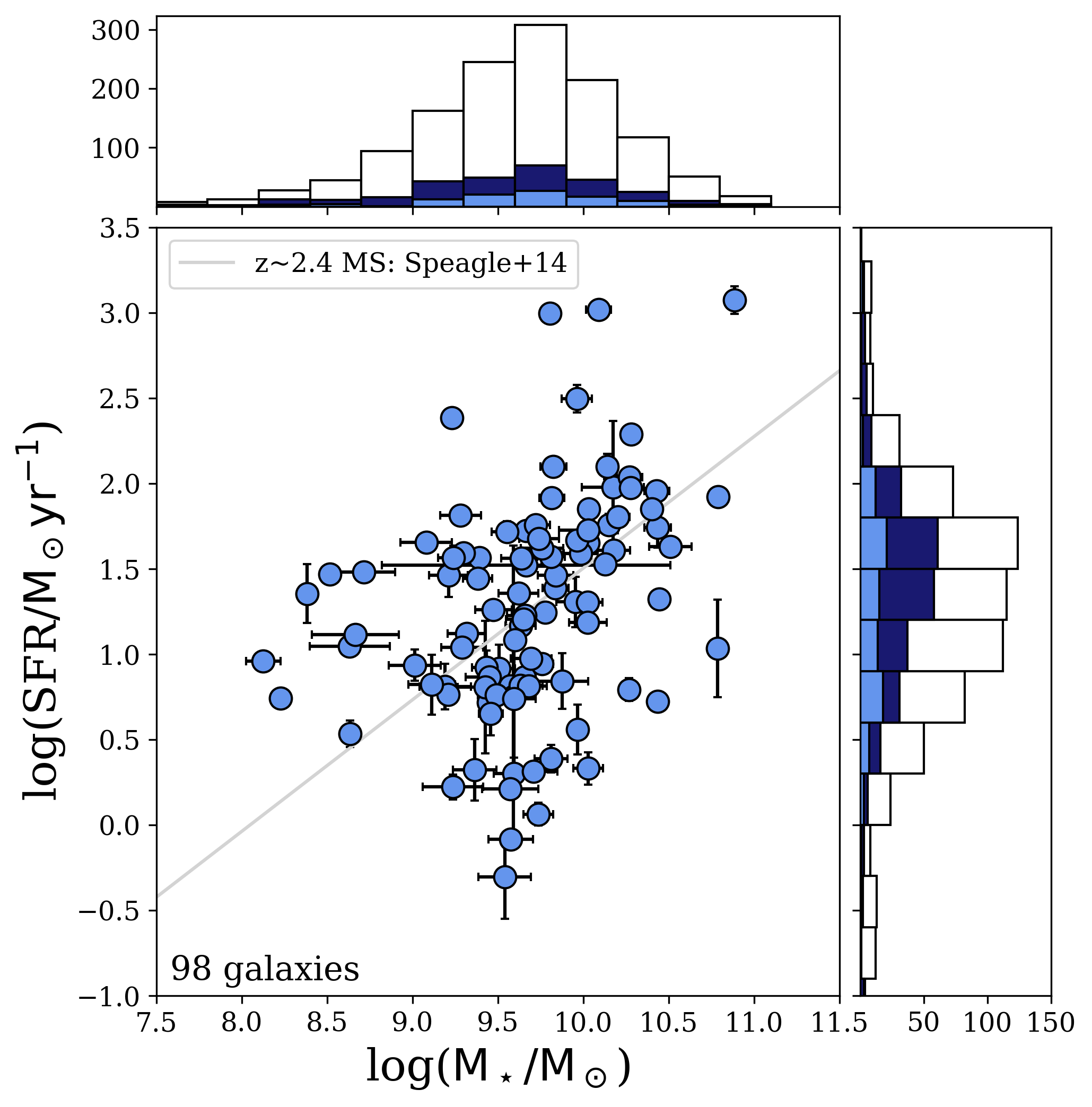}
    \caption{SFR vs.\ stellar mass (\Mstar) for our outflow sample (blue circles). Here, \Mstar\ is derived from SED fitting (Section \ref{sec:SED}) and SFR is calculated using the metallicity-independent conversion from \cite{Kennicutt12}, detailed in Section \ref{sec:SFR}. 
    The gray line represents the star-forming main sequence at redshift $z\sim2.4$ \citep{Speagle14}. The top and right histograms, respectively, represent the log(SFR) and log(\Mstar) distributions of our outflow sample (light blue) compared to the high--SNR sample (dark blue) and the parent KBSS+KLCS sample (white). KS-tests (Section \ref{sec: occurence}) show that there are no biases in stellar mass, but the high--SNR and outflow samples are biased towards higher star-forming galaxies. }
    \label{fig:SFMR}
\end{figure}

\subsection{Slit-loss and reddening}
\label{sec: reddening}
We correct the \ha\ and \hb\ line fluxes for slit losses and dust reddening. \cite{Strom17} used a Markov chain Monte Carlo method to infer the slit losses for individual galaxies across the $J$, $H$, and $K$ bands. We adopt the same methodology here. For reference, a typical slit-loss correction for the $K$ band ranges from a factor of 1.3 to 2.9.  

To account for dust attenuation, we adopt the Milky Way dust extinction curve with $R_V = 3.1$ from \cite{Cardelli89}. We estimate the dust extinction from the Balmer decrement as follows. When \ha\ and \hb\ are available, we compare their observed line flux ratio, using the total (broad + narrow) flux, to the theoretical Case-B value of 2.86 \citep{Osterbrock06}. If either line is unavailable, typically for $z>3$, we consider higher-order Balmer lines, such as H$\gamma$ and H$\delta$, when they have SNR $>$ 2. All galaxies in the outflow sample have a complete pair of Balmer lines.

We deredden \ha\ and \hb\ to derive galaxy parameters, including line luminosity and SFR. For reference, $k_{H\alpha}$ = 2.53 when adopting the \cite{Cardelli89} extinction curve. The median extinction, when using the total line profile, and color excess in our outflow sample are $A_\textrm{\ha}$ = 0.61 mag and E(B--V) = 0.24. These values are consistent with the whole KBSS sample, which has $A_\textrm{\ha}$ = 0.63 mag and E(B--V) = 0.25 \citep{Strom17}.

\subsection{Star-formation rate}
\label{sec:SFR}
For a more direct comparison to existing literature, we calculate the instantaneous SFR using the conversion from \cite{Kennicutt12}: $\textrm{SFR}(\textrm{\ha}) = \textrm{log}(L_{\textrm{\ha,N}})-\textrm{log}(C_\textrm{x})$ where $L_{\textrm{\ha,N}}$ is the dust-corrected luminosity of the narrow \ha\ component, and $C_\textrm{x}$ = 41.27, assuming the \cite{Chabrier03} IMF. 

Then, to understand the spatial concentration of star formation, we measure the SFR surface density \SigmaSFR, defined as:
\begin{equation}
    \label{eq: sigmaSFR}
    \Sigma_\mathrm{SFR} = \frac{\mathrm{SFR}}{2\pi {R_E}^2}
\end{equation}

Here, $R_E$ is the effective radius of the host galaxy, obtained using measurements from \cite{Chen2021}. Using \texttt{GALFIT 3.0} \citep{Peng2010}, they fit S\'ersic profiles convolved with point-spread functions to Hubble Space Telescope (HST) imaging or ground-based imaging from Keck/LRIS or Palomar/P200. The size measurements used in this study are derived from a mixture of rest-UV and rest-optical imaging. Most outflow studies use rest-optical imaging because they can trace the extended gas of galaxies. In this study, we prioritize space-based rest-optical measurements using the HST F140W or F160W filters when possible. However, more than half of the size measurements in the final outflow sample are derived from ground-based imaging in both the rest-optical (N=27) and rest-UV (N=17) bands; only 5 galaxies have space-based rest-UV imaging in the F814W filter.

Rest-UV size measurements are typically larger compared to those from the rest-optical. For example, \cite{vanderWel14} finds that galaxy size decreases with increasing wavelength in the 3D-HST survey; correction factors can be estimated using Equations 1 and 2 in their paper. Taking the median stellar mass (10$^{9.63}$ \Msun) and redshift ($z=2.32$) of galaxies with rest-UV imaging in our outflow sample, rest-UV sizes are only larger by a factor of 1.0077. Due to this small difference, size corrections may not be necessary.

In addition, we conduct a similar analysis as \cite{Llerena23} by comparing space-based optical (HST F140W or F160W) with ground-based rest-UV (Keck/LRIS Rs) sizes from the KBSS+KLCS parent sample. As ground-based instruments have lower spatial resolution, it is challenging to identify a significant offset or correction factor for the rest-UV sizes. To account for the wavelength dependence of size measurements and the lower spatial resolution of ground-based instruments, we adopt a systematic scatter of 0.58 arcsec in $R_E$, corresponding to the dispersion between HST rest-optical and ground-based measurements. These size measurements are not available for most of the KLCS galaxies, so we exclude them in further analyses that require $R_E$. Overall, the properties of each galaxy are presented in the Appendix. (Please find the table in the ApJ version of the paper.) 

\section{Results}
\label{sec:results}

\subsection{Outflow Occurence}
\label{sec: occurence}
Our high--SNR galaxy sample is composed of 323 KBSS+KLCS galaxies with either SNR(\ha)$>$50 or SNR([\oiii])$>$50 (Section \ref{sec:High-SNR Sample}). After our double-Gaussian fitting procedure (Section \ref{sec: gaussian}), our final outflow sample consists of 98 galaxies.

\begin{table}[h]
\centering
\caption{\textbf{Outflow Occurrence}}\label{tab:occurrences}
\begin{tabular}{lccc}
\toprule
\textbf{ } & \textbf{N$_\textrm{gal}$} & \textbf{N$_\textrm{out}$} & \textbf{Occur.} \\
\midrule
\textbf{High--SNR galaxy sample} & 323 & 98 & 30.3$\%$ \\
\hline
\textbf{[\oiii]--detected gal.} & 311 & 98 & 31.5$\%$ \\
\textbf{SNR([\oiii])$>$50 gal.} & 263 & 94 & 35.7$\%$\\
\textbf{\ha--detected gal} & 259 & 78 & 30.1$\%$\\
\textbf{SNR(\ha)$>$50 gal.} & 138 & 45 & 32.6$\%$\\
\bottomrule
\end{tabular}
\tablecomments{Outflow occurrence rates for different groups of galaxies used in this paper. The high--SNR galaxy sample is defined in Section \ref{sec:High-SNR Sample}. Among them, the galaxies with detected [\oiii] or \ha\ are shown in rows 2 and 4, respectively. Galaxies with high--SNR emission lines are shown in rows 3 and 5. The number of galaxies (N$_\textrm{gal}$) and the number of galaxies with outflows (N$_\textrm{out}$) in each group are presented in columns 2 and 3. The last column shows the outflow occurrence rate, defined by N$_\textrm{out}$/N$_\textrm{gal}$. }
\end{table}

We show the SFR–\Mstar\ relationship (SFMS) for our outflow sample in Figure~\ref{fig:SFMR} as blue circles. The light blue, dark blue, and white histograms represent the distributions of the outflow sample (N=98), the high--SNR galaxy sample (N=306), and the full KBSS+KLCS sample (N=1650), respectively. We note that 17 KBSS galaxies in the full sample do not have accurate mass measurements due to poor photometry. 

To test whether our selection criteria introduce a bias towards brighter and more highly star-forming galaxies, we perform the Kolmogorov–Smirnov (KS) test between the full, parent, and outflow samples. If the KS p-value is less than 0.05, then the two samples are statistically drawn from different distributions. The KS p-values for stellar mass are 0.88 between the parent and high--SNR samples and 0.34 between the high--SNR and outflow samples. For SFR, the KS p-values are 0.02 between the parent and high--SNR samples and 0.48 between the high--SNR and outflow samples. Therefore, these tests suggest that there are no significant biases in the mass distributions for the samples. However, there is a bias toward higher-star-forming galaxies in both our high--SNR and outflow samples. This bias is expected because we prioritize galaxies with higher SNR, so fainter, lower star-forming galaxies are less likely to be included in the final selection. Figure \ref{fig:sfr vs snr} shows the relationship between SFR and SNR. Although our sample is more complete for SFR$>$1 \Msun yr$^{-1}$, excluding galaxies with lower SFR does not affect our main conclusions.

The outflow occurrence rate for different groups of galaxies defined in this paper is listed in Table \ref{tab:occurrences}. The number of galaxies and galaxies with outflows are listed in the second and third columns, respectively. The corresponding outflow occurrence rates are shown in the fourth column and range from 30--35$\%$. It is worth noting that our outflow sample size of 98 is two to three times larger than that in previous literature.
 
Outflow detection rates vary widely across the literature, but our detection rate falls within the range of previously reported rates, given our differences in sample size. For instance, \cite{Weldon24} detected outflows in 10$\%$ of 598 galaxies; \cite{Llerena23} detected outflows in 65$\%$ of 35 galaxies. In addition, our detection rate is consistent with higher redshift samples, where \cite{XuY25} and \cite{Carniani24} detect outflows in 30$\%$ of 130 galaxies and 25--40$\%$ of 52 galaxies, respectively. 

We note that the outflow detection rate increases with SNR and spectral resolution, as discussed by \cite{Freeman19} and \cite{Xu25}. It is also sensitive to the time variability of outflows and outflow geometry (e.g., whether the galaxy is face-on or edge-on), owing to the projection of velocities along our line of sight. We investigate the latter point further in Section \ref{sec: detectability}.

\begin{table*}[ht!]
\centering
\caption{Summary of Literature Studies Used for Comparisons}
\label{tab:literature}
\begin{tabular}{l l l l l l}
\hline\hline
Reference & Telescope/Instrument & $z$ & $R$ & $\#$ & Diagnostics \\
\hline
This work & Keck/MOSFIRE & 1.5--3.3 & 3300--3600 & 98& H$\alpha$, [\oiii] $ \lambda\lambda$4959, 5007 \\
\hline
\cite{Marasco23} & VLT/MUSE & $\sim$0 & 1800--3600 & 19 & H$\alpha$ \\
\cite{Mcquinn2019} & KPNO$^{(1)}$/Mayall, KPNO/Bok & $\sim$0 & $-^{(2)}$ & 12 & H$\alpha$ \\
\cite{Reichardt24} & Keck/KCWI & $\sim$0 & 2500 & 10 & H$\beta$, [\oiii] $\lambda$5007 \\
\cite{Xu25} & Keck/ESI, VLT/X-Shooter & $\sim$0 & 4000--5000 & 33 & H$\alpha$, [\oiii] $\lambda$5007, [\oiii]$ \lambda$4363 \\
\cite{Forster19} & VLT/KMOS & 0.6--2.7 & 3600--4200 & 49$^{(2)}$ & H$\alpha$ [N\,II]$\lambda\lambda$6548, 6584 \\
\cite{Davies19} & VLT/SINFONI & 2--2.6 & 3500 & 28$^{(3)}$ & H$\alpha$ \\
\cite{Llerena23} & Keck/MOSFIRE, VLT/X-Shooter & $\sim$3 & 3300--5600 & 23 & [\oiii] $\lambda\lambda$4959, 5007 \\
\cite{Weldon24} & Keck/MOSFIRE & 2--3 & 3300--3600 & 33$^{(4)}$ & H$\alpha$, [\oiii] $\lambda\lambda$4959, 5007 \\
\cite{Cooper25} & JWST/NIRSpec & 2.5--9 & 1000 & 34 & [\oiii] $\lambda$5007 \\
\cite{XuY25} & 
JWST/NIRSpec
& 3--9 & 2700 & 30 &  H$\alpha$, [\oiii] $\lambda\lambda$4959, 5007 \\
\cite{Carniani24} & JWST/NIRSpec & 3--9 & 2700 & 13 & H$\alpha$, [\oiii] $\lambda\lambda$4959, 5007 \\
\cite{Saldana-Lopez25} & JWST/NIRSpec & 4--8 & 2700 & 5 & H$\alpha$, [\oiii] $\lambda\lambda$4959, 5007 \\
\hline

\end{tabular}
\tablecomments{\textbf{$^{(1)}$} Kitt Peak National Observatory \textbf{$^{(2)}$} \cite{Mcquinn2019} adopts narrow band imaging of \ha\ to probe the existence of outflows.\textbf{$^{(2,5)}$} \cite{Forster19} creates a single stack using the spectra of 49 galaxies from $z = 1.7-2.7$.\textbf{$^{(3)}$} \cite{Davies19} creates 5 stacks using the spectra of 28 galaxies, binned by \SigmaSFR. \textbf{$^{(4)}$} Only including the outflow measurements made using \ha\ from \cite{Weldon24}.
}
\end{table*}

\subsection{Outflow velocity}
\label{sec:outflow velocity}
Outflow velocity is one of the most foundational quantities for describing warm-ionized outflows. Following previous studies \citep[e.g.][]{Llerena23}, we define the maximum outflow velocity as
\begin{equation}
    \label{eq: vmax}
    V_\textrm{max} = |\Delta v| + 2\sigma_{B}
\end{equation}
where $\Delta v$ is the difference in velocity centers between the broad and narrow components, and $\sigma_{B}$ is the velocity dispersion of the broad component. This $V_\textrm{max}$ is approximately the 98th percentile of outflowing gas velocity. The measured velocity of the sample is approximately \Vmax\ $\approx$ 200--800 \kms, consistent with typical KBSS outflow velocities (500--800 \kms) measured using absorption lines \citep[][and Theios et al. (in prep)]{Steidel10, Trainor15}.

\begin{figure*}[h]
    \centering
    \includegraphics[width=0.94\linewidth]{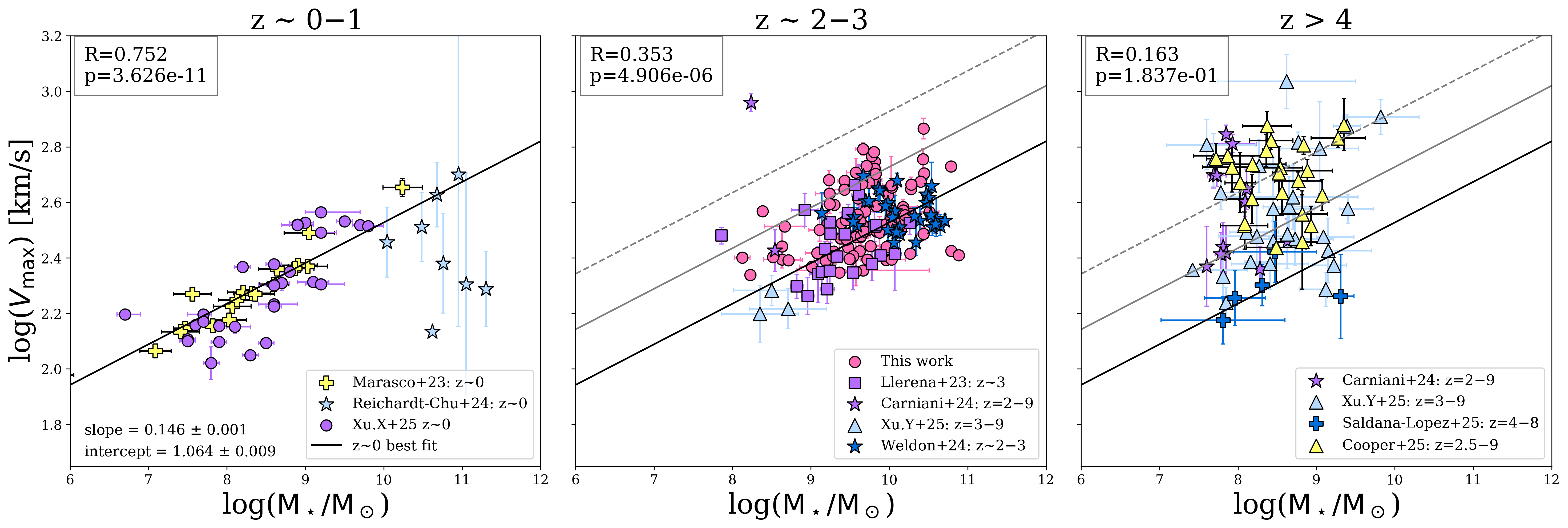} %
    \includegraphics[width=0.94\linewidth]{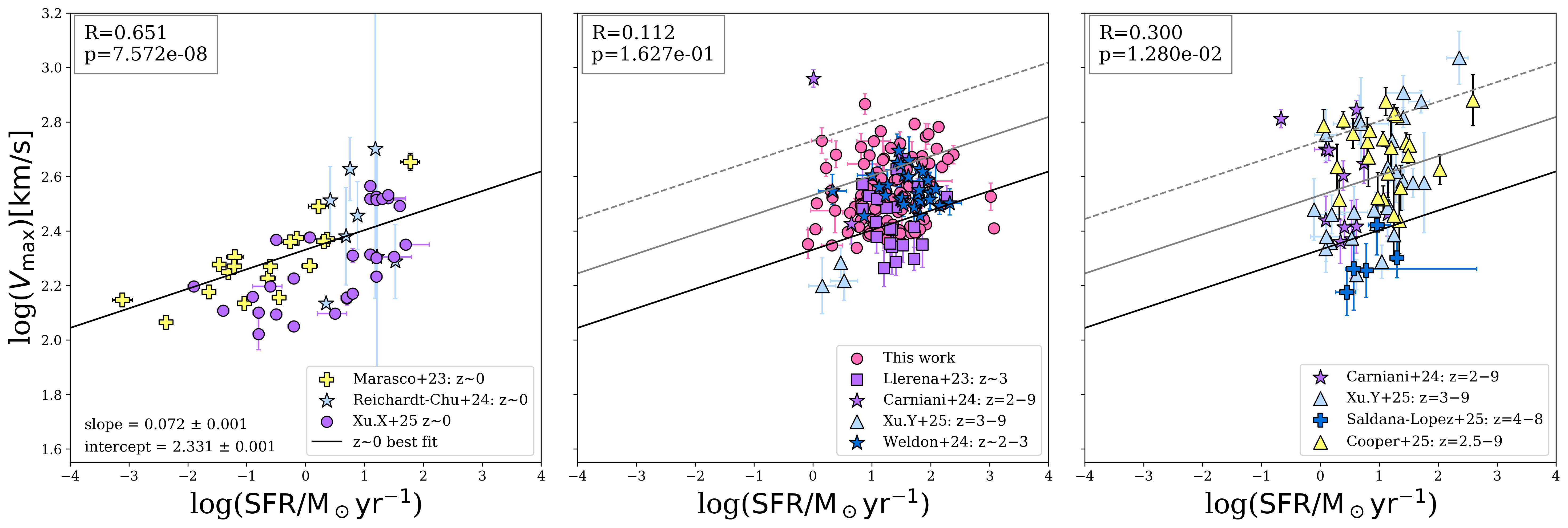} %
    \includegraphics[width=0.94\linewidth]{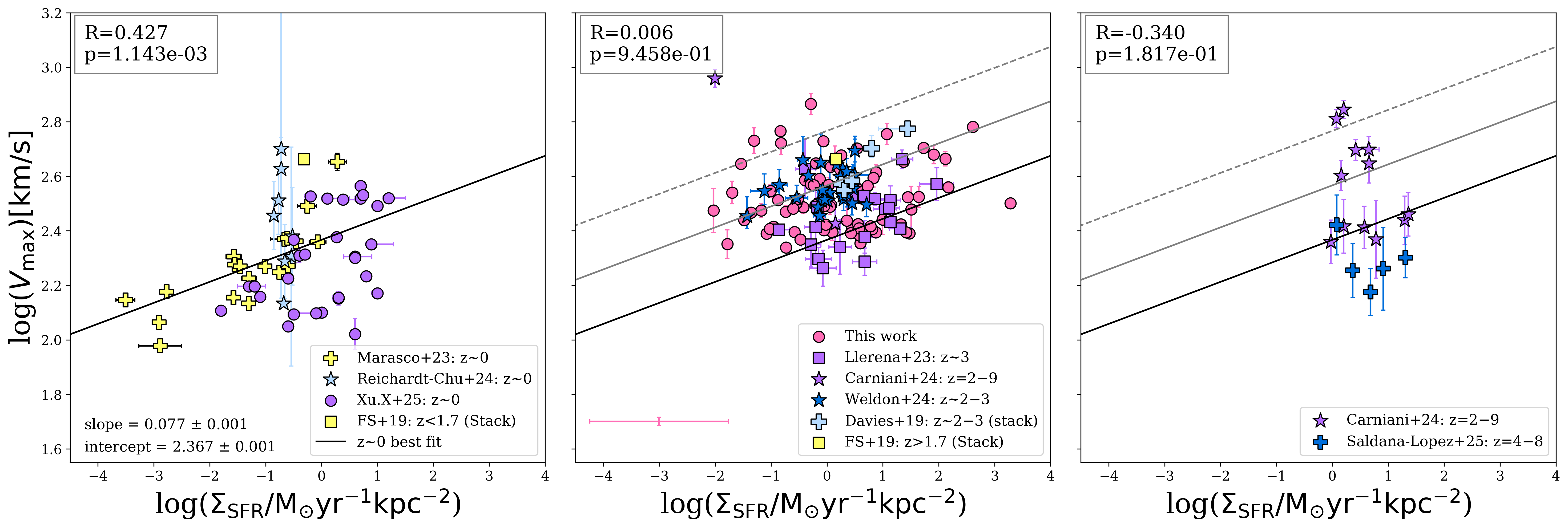}
    \caption{log(\Vmax) as a function of log(\Mstar) (top), SFR (middle), and  \SigmaSFR\ (bottom) over three redshift ranges: $z \sim 0-1$ (left),  $z \sim 2-3$ (middle), and $z \gtrsim 4$ (right). For all panels, the best-fit line for the local samples is shown by the solid black line. The solid and dashed gray lines represent vertical shifts of the best fit by 0.2 and 0.4 dex, respectively. In the bottom middle panel, the magenta cross indicates the median uncertainties in log(\SigmaSFR) and log(\Vmax), with values of 1.24 and 0.016 dex, respectively. Samples at $z \sim 0-1$ (e.g. \cite{Forster19} (denoted as ``FS+19"; yellow square), \cite{Marasco23} (yellow crosses),  \cite{Reichardt24} (light blue stars), and \cite{Xu25}(purple circles)) show significant correlation between galaxy and outflow properties. At $z \sim 2-3$, our outflow sample is denoted by the magenta circle. We compare our outflow sample to other Cosmic Noon studies: \cite{Davies19} (light blue crosses), \cite{Forster19} (yellow square), \cite{Llerena23} (purple squares), \cite{Weldon24}(dark blue stars), \cite{Carniani24} (purple stars), and \cite{XuY25} (light blue triangles). The latter two studies also have samples at $z \gtrsim 4$. Additionally, we include \cite{Saldana-Lopez25}(dark blue crosses) and \cite{Cooper25} (yellow triangles) for the sample at $z \gtrsim 4$. For literature samples, we include uncertainties when available. The best-fit trend at local redshifts describes a large fraction of galaxies at Cosmic Noon and higher redshifts. However, Cosmic Noon galaxies tend to host higher velocity outflows for fixed \Mstar, SFR, and \SigmaSFR.} 
    \label{fig:vmax scaling}
\end{figure*}

\begin{figure*}
\centering
\includegraphics[width=1\linewidth]{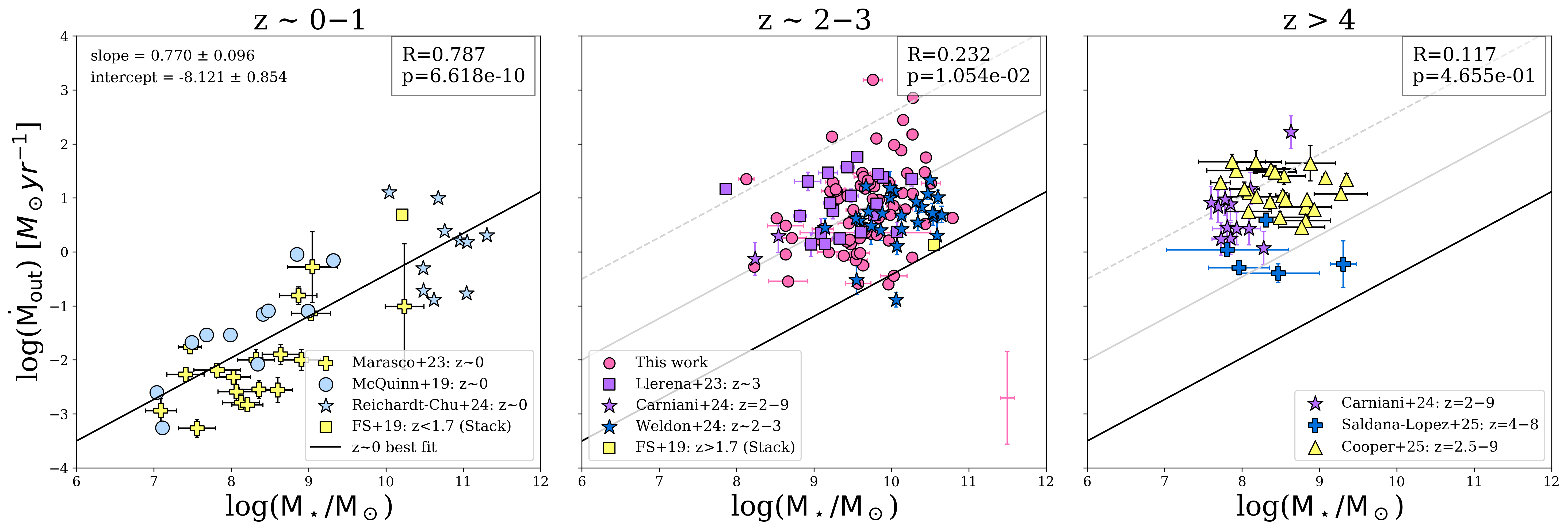}
\caption{\MdotOut\ as a function of log(\Mstar) across redshift ranges. We compare the same studies as in Figure \ref{fig:vmax scaling}, in addition to \cite{Mcquinn2019} (blue circle) at low redshift, and the mass stack from \cite{Forster19} (yellow square) at low redshift and Cosmic Noon. The best-fit line at $z \sim 0-1$ is shown in the three panels as a solid black line. The gray solid and dashed lines in the middle and right panels represent a vertical shift of the local best-fit line by 1.5 and 3 dex, respectively. Galaxies at Cosmic Noon tend to host stronger outflows than local galaxies in the same mass range (Section \ref{sec:redshift evo}).}
\label{fig:mdot vs mass}
\end{figure*}

To evaluate the outflow scaling relations and their possible evolution over cosmic time, we contrast log($V_\textrm{max}$) with various galaxy properties ($M_\star$, SFR, and \SigmaSFR) at low-$z$ ($z<1$), medium-$z$ ($z\sim2-3$), and high-$z$ ($z>4$). In Figure \ref{fig:vmax scaling}, our galaxies are shown in magenta, and we overlay the results from literature studies in other colors. All of the studies shown in Table \ref{tab:literature} select their samples using double-Gaussian fitting to identify broad emission-line components. The details of these studies are summarized in Table \ref{tab:literature}. We note that some previous works define the outflow velocity as $|\Delta v| + \textrm{FWHM}_\textrm{B}/2$ \citep[e.g.][]{Marasco23, XuY25}. For these works, we recalculated their \Vmax\ to be consistent with our definition in Equation \ref{eq: vmax}. 

The leftmost panels show strong, significant correlations in the collective low-$z$ ($z<1$) outflow studies. We fit a linear line using \texttt{scipy.curve$\_$fit} (shown in black) to all the low-$z$ literature samples collectively to serve as a baseline for comparison at all redshifts. In the middle panel, we find a moderately strong correlation between log(\Vmax) and log(\Mstar), when combining the results from multiple Cosmic Noon studies ($R=0.353$, $p=4.906 \times 10^{-6}$). This showcases the importance of having a large sample of individual galaxies that spans a range of galaxy properties. For SFR and \SigmaSFR, the correlations of the total samples at Cosmic Noon and beyond (middle and right columns) with \Vmax\ are weak to moderate and statistically less significant. 

The absence of expected trends at higher redshifts might be explained by the smaller dynamic ranges in the galaxy properties typically of interest (e.g., \Mstar\ and SFR). Observations at these redshifts are biased towards brighter galaxies with higher SFR due to the difficulty of detecting outflows in faint, distant galaxies. We notice an increasing upward scatter in \Vmax\ with redshift, which might suggest that some outflows are intrinsically stronger in the medium to high-$z$ Universe. To test this idea, we overlay the best-fit line from low-$z$ on the right two columns (solid black line), and shift it vertically by 0.2 and 0.4 dex (gray solid and dashed lines). We find that while the low-$z$ trends can describe a large population of $z>2$ galaxies with relatively mild outflow velocities, other galaxies tend to have larger \Vmax. Additionally, weak correlations are seen in the \SigmaSFR\ v.s. \Vmax\ for both intermediate- and high-redshift samples (bottom-middle and bottom-right panels). At $z > 4$, there are far fewer galaxies present to establish a relation. Moreover, the ambiguity at $z > 2$ is likely due to the difficulty of extracting spatial information for measuring $R_E$; the large uncertainties in $R_E$ likely contribute to the scatter in the \SigmaSFR\ vs. \Vmax\ relation.

\subsection{Mass outflow rate}
The mass outflow rate (\MdotOut) measures the rate at which mass is expelled from the host galaxy. We adopt a simple outflow model from \cite{Newman12a} and \cite{Genzel11}, assuming that outflows have a constant radial velocity and a constant \MdotOut. Additionally, they assume Case B recombination and that all the broad component gas is photoionized at a temperature of $T$ = $10^4$ K. Following the simplified functional form from \cite{Weldon24}, the mass outflow rate can be calculated as
\begin{equation}
    \label{eq: Mdot}
    \dot{M}_\mathrm{out} = \frac{1.36 m_\textrm{H}}{\gamma_{\textrm{H}\alpha}n_e} L_{\mathrm{H}\alpha, b} \frac{V_\mathrm{max}}{R_E} \ .
\end{equation}

Here, $m_\textrm{H}$ is the mass of hydrogen in grams, $\gamma_{H\alpha}$ = 3.56 $\times$ $10^{-25} T_e^{-0.91}$ erg cm$^{-3}$ s$^{-1}$ is the emissivity of H$\alpha$ at an electron temperature of $T_e =10^4$ K, and $L_{\textrm{H}\alpha,b}$ is the dust-corrected luminosity from the broad \ha\ component. We assume an electron density of $n_e = 300\pm 200$ cm$^{-3}$, consistent with the range of electron densities $n_e \sim 224 - 324$ cm$^{-3}$ of typical H II regions measured using [\oii] $\lambda\lambda$3726, 3729 in stacked spectra of KBSS galaxies \citep{Strom17, Sanders2016}. 

The assumption that the electron density of outflows is the same as that of typical H II regions at Cosmic Noon is likely a lower limit. In local galaxies, the outflow electron density has been found to be at most 3 times higher than the galactic disk density \citep{Fluetsch21, Marasco23}. However, the outflow electron densities at higher redshifts are more difficult to constrain and contribute large uncertainties to \MdotOut\ and $\eta_m$ measurements. Cosmic Noon studies typically assume a range of $n_e = 200-600$ cm$^{-3}$ \citep{Davies19, Llerena23, Weldon24}. Our assumption of $n_e= 300$ cm$^{-3}$ is consistent with previous literature, but it is noted that if we assume a higher density of, e.g., $n_e = 600$ cm$^{-3}$, our analytic measurements for \MdotOut\ will systematically shift downwards by a factor of two.

In Figure \ref{fig:mdot vs mass}, we find that \MdotOut\ is strongly correlated with log(\Mstar) in the local Universe ($R=0.787$ and $p=6.62\times10^{-10}$) and moderately correlated at Cosmic Noon ($R=0.232$ and $p=1.05\times10^{-2}$). The median uncertainty on log(\MdotOut) is $\pm$0.862 dex; however, the true uncertainties on \MdotOut\ are likely larger due to the uncertainties on galaxy sizes and electron densities. It is also obvious that the Cosmic Noon galaxies have higher \MdotOut\ than local galaxies at fixed \Mstar\ by 1.5--3 dex. Trends at z $\gtrsim$\ 4 are insignificant, although this might be attributed to the much smaller dynamic range. We discuss the difference between outflows at Cosmic Noon and the local Universe in more detail in Section \ref{sec:redshift evo}.

\begin{figure*}
    \centering
    \includegraphics[width=1\linewidth]{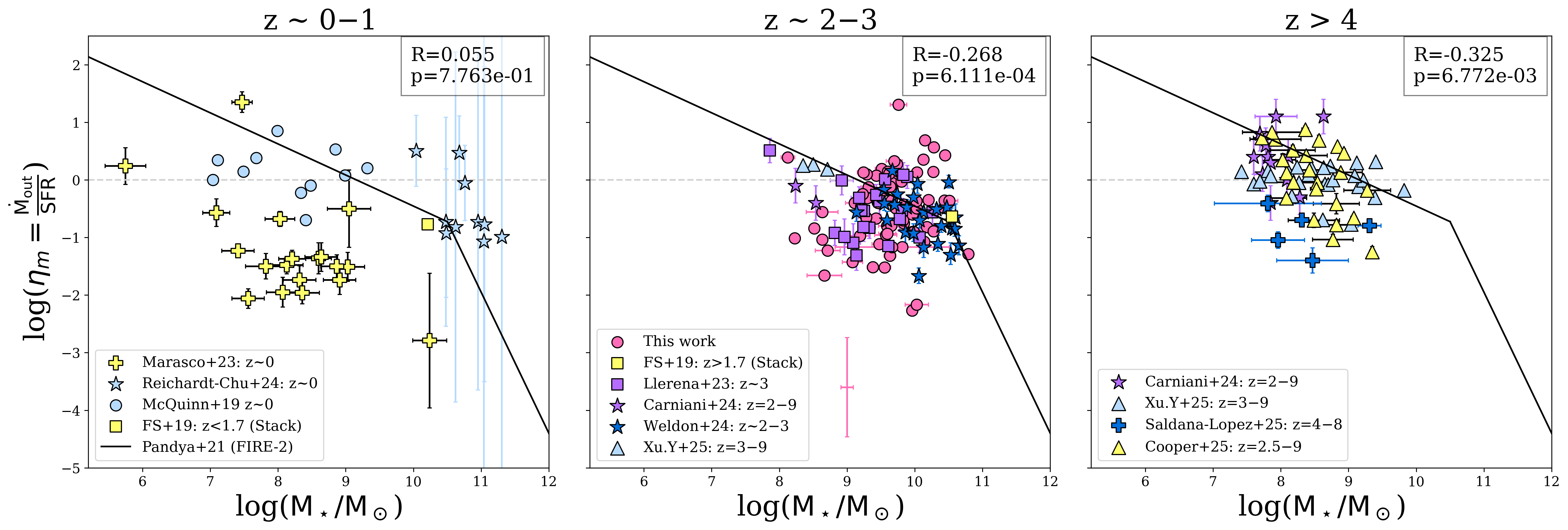} 
    \caption{Warm-phase mass loading factor ($\eta_m$) as a function of log(\Mstar) throughout multiple redshifts. The samples and markers are the same as in Figure \ref{fig:vmax scaling} and Figure \ref{fig:mdot vs mass}. The magenta cross in the middle panel represents the median uncertainty on log(\Mstar/\Msun) and log(\MdotOut). We compare all samples to the theoretical warm-phase mass loading factor predicted from the FIRE-2 simulation in the black solid line, measured at 0.1--0.2 $\textrm{R}_{\textrm{vir}}$ \citep{Pandya21}. The gray dashed line represents when \MdotOut=SFR. Galaxies across all redshift ranges show the expected negative trend between $\eta_m$ and log(\Mstar), as outflows in more massive galaxies are likely confined by gravity. We note that \cite{Marasco23} (yellow crosses) measure a lower $\eta_m$ than other studies, but still follows a negative trend. There appears to be little redshift evolution across all samples, suggesting that the suppression of stellar feedback at high dark matter halo masses ($\approx$ $10^{12}$ \Msun) is relatively independent of redshift.}
    \label{fig: eta vs mass}
\end{figure*}

\subsection{Mass loading factor}
\label{sec:massloading}
The mass loading factor $\eta_m$ is a measure of how efficiently an outflowing gas removes material from its host galaxy with respect to star formation, and is defined as the ratio of the mass outflow rate to SFR \citep[e.g.,][]{Veilleux05}: 

\begin{equation}
    \label{eq:eta}
    \eta_m = \frac{\dot{M}_\mathrm{out}}{\mathrm{SFR}} . 
\end{equation}

In Figure \ref{fig: eta vs mass}, we show that the combined samples across the three redshift ranges likely follow an expected negative trend between $\eta_m$ and log(\Mstar). We note that there is an offset between \cite{Marasco23} (yellow cross) and other $z<1$ samples, which is described in more detail in Section 5.2 of their paper. To briefly summarize, \cite{Marasco23} attributes this offset to line-of-sight projection effects or to the warm-phase gas making up less than 1$\%$ of the total outflowing wind mass.

Samples at $z<1$ show little to no correlation with $R=0.055$. The correlation becomes moderate for higher redshifts, with $R=-0.268$ and $p=6.111\times 10^{-4}$ for $z\sim 2-3$, and $R=-0.325$ and $p=6.772 \times 10^{-3}$ for $z>4$. The median uncertainty on log($\eta_m$) is $\pm$0.862 dex. Overall, the negative correlation suggests that galactic outflows expel gas more efficiently in lower-mass galaxies. Our collective samples also agree with the theoretical predictions from the FIRE-2 simulated mass loading factor up to log(\Mstar/\Msun) $\lesssim$ 10.5 \citep[black solid line, taken from][]{Pandya21}. However, \cite{Pandya21} measures $\eta_m$ at 0.1--0.2 $\textrm{R}_\textrm{vir}$, which is a larger distance than the median effective radius ($\approx 0.015\ \textrm{R}_\textrm{vir}$) in this study.

\begin{figure*}
    \centering
    \includegraphics[width=1\linewidth]{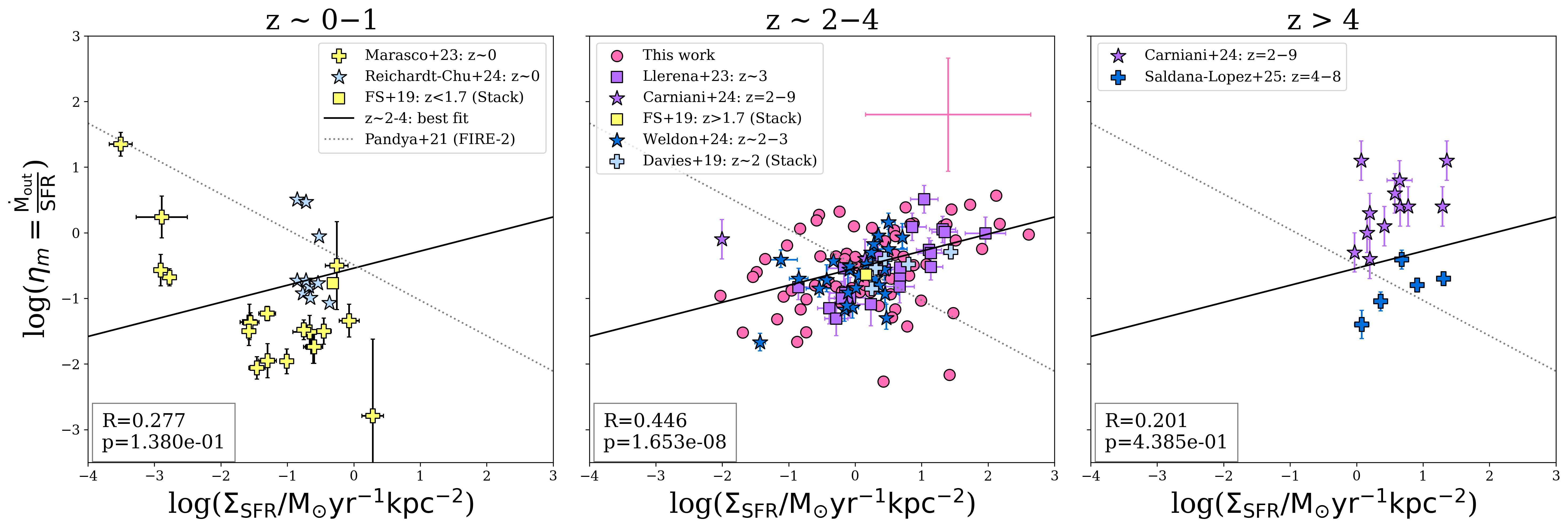}
    \caption{Mass loading factor as a function of \SigmaSFR\ throughout multiple redshift ranges. Samples and markers are the same as in Figure \ref{fig: eta vs mass} with the addition of \cite{Davies19} at the intermediate redshifts. The solid black line represents the best fit at $z \gtrsim 2-3$, and the gray dotted line represents the FIRE-2 prediction from \cite{Pandya21}. Contrary to theoretical predictions, observations of galaxies at Cosmic Noon suggest a strong positive correlation between $\eta$ and \SigmaSFR\ (see Section \ref{sec:massloading}). }
    \label{fig:eta vs sigmaSFR}
\end{figure*}

The sharp decrease of $\eta_m$ at $\approx 10^{10.5}$ \Msun\ in the FIRE-2 simulation is attributed to gravitational and pressure confinement \citep[][]{Stern2021, Byrne23, Hopkins2023}. Outflows from massive galaxies are less likely to reach escape velocity and get expelled from their host. Furthermore, the inner CGM virializes when \Mstar/\Msun $\gtrsim 10^{10.5}$, corresponding to a halo mass of $\textrm{M}_\textrm{halo}$/\Msun$\sim 10^{12}$. When this happens, the thermal pressure in the inner CGM increases, potentially playing a role in confining stellar feedback to weak fountains at the disk-halo interface. In addition, most of the gas that leaves the galaxies remains hot, making warm outflows rare in FIRE-2 simulations when \Mstar/\Msun\ $\gtrsim 10^{10.5}$ \Msun\ \citep{Pandya21}.

To further understand the efficiency of outflows, we look at the relationship between $\eta_m$ and \SigmaSFR\ in Figure \ref{fig:eta vs sigmaSFR}. No clear correlations are observed in the $z<1$ and $z>4$ samples, likely due to their small dynamic range in \SigmaSFR. However, we do see significant correlations at Cosmic Noon. Prior to adding our outflow sample, the collective $z\sim2-3$ literature samples show a moderate positive correlation with $R=0.409$ and $p=2.4\times10^{-3}$. After adding our outflow sample, the correlation strength and significance increased to $R=0.563$ and $p=1.4\times10^{-14}$. 

Naively, a positive trend is expected because, if SFR is concentrated, outflows are likely to experience greater pressure to escape the gravitational well of their host galaxies \citep{Newman12a}. Moreover, local outflows are a consequence of high \SigmaSFR\ \citep{Chen10}. Interestingly, the observed trend at Cosmic Noon is different from the FIRE-2 predictions, which show a negative correlation \citep[gray dotted line][]{Pandya21}. This may be because the time-dependent characteristic of both $\eta_m$ and \SigmaSFR. In simulations, there is a time delay between when star formation and outflows are measured \citep[see Figure 2 in][]{Muratov15}. Therefore, outflows are measured after the burst of star formation happens for 10s--100s of Myr, depending on the size of the galaxy and the velocity of the outflow. In contrast, our observations probe the current star formation from \ha\ ($\sim$ 10 Myr) and the current outflow properties. Given these differences, it is difficult to make a direct comparison between observations and simulations in Figure \ref{fig:eta vs sigmaSFR} since we cannot trace back the exact burst of star formation that caused the observed outflows. 

\section{Discussion} 
\label{sec:discuss}
\subsection{Bound vs. Unbound Outflows}
\label{sec:bound unbound}
To understand how outflows influence galactic feedback, it is important to examine whether the gas is expelled into the CGM or IGM, or recycled back into the host galaxy. We compare the escape velocity $V_\textrm{esc}$ of the host galaxies to \Vmax\ as a function of the dynamical mass $M_\textrm{dyn}$ of galaxies. We adopt the functional form of $V_\textrm{esc}$ from \cite{Arribas2014}: 
\begin{equation}
    \label{eq:Vesc}
    V_\mathrm{esc} \approx\!\left(\frac{2M_\mathrm{dyn}G(1+\mathrm{ln}(\mathrm{R}_\mathrm{vir}/r))}{3r}\right)^{1/2} .
\end{equation}

We define $r=R_E$, R$_\textrm{vir}$ is the average virial radius of the outflow sample, and $G$ is the gravitational constant. Taking the median stellar mass of the outflow sample (\Mstar = 10$^{9.7}$ \Msun), the corresponding mass of the dark matter halo is M$_\textrm{halo}$ = 10$^{11.8}$ \Msun (see Figure 12 of \citealt{Shuntov2022}). The resulting virial radius is then R$_\textrm{vir}$ = 74 kpc. Additionally, \cite{Erb2006} defines $M_\textrm{dyn}$ as:
\begin{equation}
    \label{eq:Mdyn}
    M_\textrm{dyn} \approx \frac{C\sigma^2R_E}{G} \ ,
\end{equation}
where C is a constant factor that depends on the galaxy's mass and velocity distributions, and $\sigma$ is the velocity dispersion of the narrow emission line component. Here, we assume C = 3.4 based on star-forming galaxies at $z\sim2$ in the \cite{Erb2006} sample.

\begin{figure}[]
    \centering
    \includegraphics[width=1\linewidth]{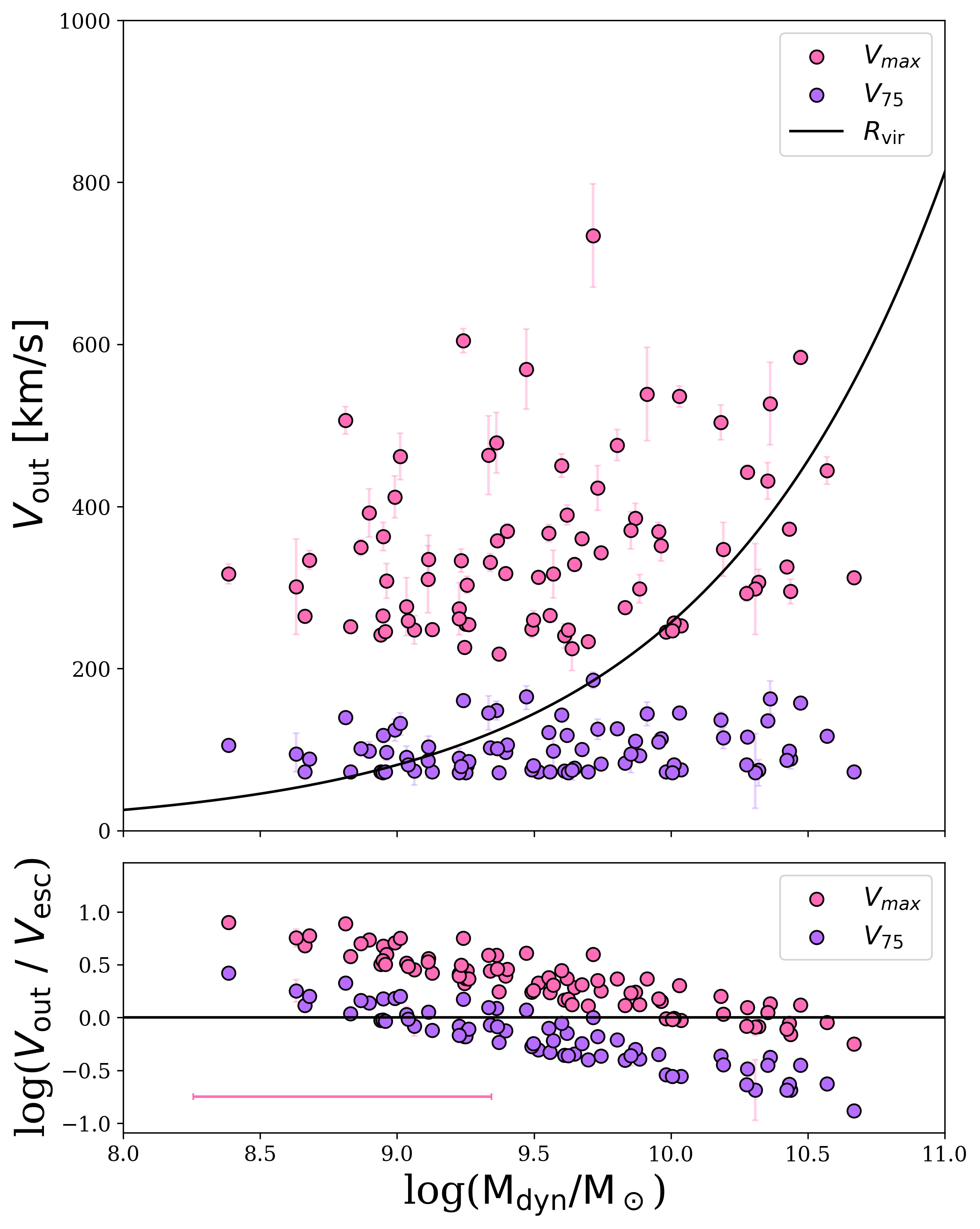} 
    \caption{Top: Outflow velocity as a function of dynamical mass. Magenta circles represent \Vmax\ and the purple circles represent $V_{75}$, the 75th percentile of outflow velocities. The solid line is an analytic expression from \cite{Arribas2014} for $V_\textrm{esc}$ measured at R$_\textrm{vir}$. Bottom: Escaped outflow velocity fraction with respect to dynamical mass. The black solid line marks the order of unity for $V_\textrm{esc}$ measured at R$_\textrm{vir}$. The horizontal magenta line represents the median uncertainty of log(M$_\textrm{dyn}$/\Msun), $\pm$ 0.544 dex. Vertical error bars are present in the figure but are small, given the log scale. The majority of detected outflowing material is traveling below $V_{75}$ and does not exceed $V_\textrm{esc}$. This material is likely recycled back into the central galaxy.}
    \label{fig:Vfrac vs Mdyn}
\end{figure}

When considering the outflow velocity, it is important to note that the majority of the outflowing gas traced by nebular emission is traveling slower than \Vmax. Therefore, we also compute the 75th percentile of outflow velocities ($V_{75}$) to determine the fraction of outflows that escape their hosts. $V_{75}$ is defined as the velocity at which the cumulative flux of the broad component is 75$\%$ of the total broad component flux. 

Figure 7 compares the outflow velocity with the host's analytic escape velocity for R$_\textrm{vir}$ = 74 kpc and r = 2 kpc, where 2 kpc is slightly larger than the median effective radius of the sample. In this case, for our outflow sample, 71.4$\%$ of the fastest outflows (i.e., with \Vmax) and 19.4$\%$ of the 75th percentile outflows (i.e., with $V_{75}$) would be able to travel beyond R$_\textrm{vir}$. Since most outflows are below V$_{75}$, they are likely to be recycled. Our result aligns with previous KBSS findings \citep{Chen2020, Prusinski25}, which used circumgalactic HI to trace gas kinematics in the CGM. They found that most neutral gas within 100 kpc of the galactic center is unlikely to escape the host potential. Though the fastest initial line-of-sight velocity ($\approx$ 600 \kms) may reach escape velocity for a $ 10^{12}$ \Msun\ dark matter halo, it is not guaranteed that all the gas will escape the central galaxy.

It is important to consider the effects that gas recycling has on the mass evolution of galaxies and dark matter halos. Observations and cosmological simulations (e.g., FIRE) of medium- to high-$z$ star-forming galaxies show that multiple cycles of starbursts can occur \citep[e.g.,][]{Marszewski25}. Although the majority of outflowing gas is retained within R$_\textrm{vir}$ in the epoch of observation (i.e., the current cycle), galaxies can eventually lose large amounts of gas after multiple cycles. For instance, retaining 75$\%$ of gas per cycle leaves only 24$\%$ after five cycles (0.75$^5$). This is consistent with the low baryon retention efficiency observed in low-mass dark matter halos, whose baryon fractions lie well below the cosmic value \citep[e.g.,][]{Behroozi19}.

\subsection{Detectability of Outflows}
\label{sec: detectability}
The true detection rates of starburst-driven outflows are difficult to constrain, as observations are limited by uncertainties in the projection effects of outflow velocities (i.e., how outflows are oriented with respect to our line of sight) and the time variability of star formation. In this subsection, we first discuss the viewing angle of our outflow sample. Then, we investigate the duty cycle and outflow geometry of these galaxies. 

In the case of disk-like galaxies, the inclination \textit{i} of the galaxies can be calculated as 
\begin{equation}
    \cos^2i = \frac{(b/a)^2 - c^2}{1-c^2}
\end{equation}
where $b/a$ is the observed axis ratio and $c$ is the major/minor axis ratio. We refer the reader to Figure 8 of \cite{Hubble1926} for an illustration of this galaxy geometry. For an edge-on thin-disk approximation, $c$ = 0. Here, we take the inclination $i= 0^\circ$ as a face-on view of the galaxy and $i=90^\circ$ as edge-on. We obtain $b/a$ from the S\'ersic profile measurements from \cite{Chen2021}. The value of $c$ for star-forming galaxies above $z > 1.5$ varies due to an increase in vertical velocity dispersion \citep[e.g.,][]{Law2009}. Using integral field spectroscopy, \cite{Genzel2008} measured $c$ from velocity maps of five $z \sim 2$ massive star-forming galaxies (log$(M_\star/M_\odot) \sim 11$) from the SINS Survey. They found that the median thickness of galaxies at $z\sim2$ is $c \approx 0.34$. Recent work by \cite{Pandya2024} measures the morphology of 50 JWST-CEERS galaxies of mass range log$(M_\star/M_\odot)$ $\sim 9-10.5$ at redshift $z \sim 0.5-8$. They find that $c \approx 0.25$ with little evolution throughout redshift. Taking $c \approx 0.25$ from \cite{Pandya2024}, we find the median inclination of our galaxies to be $i\approx 67.8^\circ$, meaning that we are not measuring the full, down-the-barrel outflow velocities; true outflow velocities can be at least two times higher (1/cos$(i)\approx 2.7$). Thus, a further exploration of outflow geometry is needed to understand why we detect fast outflows at relatively high inclinations and why outflows are prominent at Cosmic Noon.
We can estimate the full outflow occurrence rate $f_\textrm{outflow}$ at Cosmic Noon by considering our incidence rate $f_\textrm{broad}$ and the opening angle $\theta$ out of the outflow. \cite{XuY25} defines the duty cycle, or the fraction of outflows given the current epoch of star formation, as
\begin{equation}
    f_\textrm{broad} = f_\textrm{outflow} \times \frac{\Omega}{4\pi}
    \label{eq:dutycycle}
\end{equation}
where $\Omega$ is the spherical solid angle of the outflow. We will assume that the outflow is bi-conical, such that Equation \ref{eq:dutycycle} reduces to $f_\textrm{broad} = f_\textrm{outflow} \times (1-\textrm{cos}(\theta))$. Moreover, they show a degeneracy between the opening angle and the outflow occurrence rate; thus, our opening-angle estimate here is a lower limit. Using the simplified form of Equation \ref{eq:dutycycle}, our outflow incidence rate of $f_\textrm{broad} = 30 \%$ corresponds to an outflow opening angle of $\theta \gtrsim 42^\circ$ and a total outflow occurrence rate of at least $f_\textrm{outflow} \gtrsim 30 \%$. Our result is consistent with the findings in \cite{XuY25}, where they find $\theta \gtrsim 45^\circ$ and $f_\textrm{outflow} \gtrsim 30 \%$ (see Figure 11 in their paper).

While we assume the outflow geometry is bi-conical and perpendicular to the disk, it is important to note that this configuration is likely uncommon. For example, S\'ersic fits suggest that most galaxies in the KBSS sample are triaxial, and well-defined disks are rare \citep{Law2012}. Moreover, \cite{Law2012b} finds no correlation between $b/a$ and outflow properties, suggesting that galaxy inclination has little effect on the observed gas kinematics and that biconical outflows are unlikely in this sample. This may explain why we observe a high \Vmax\ despite a high average inclination of $i\approx 67.8^\circ$. If a majority of galaxies at Cosmic Noon are similar to the KBSS and KLCS samples and have irregular morphology and outflow projections, then the line-of-sight velocity is not dependent on disk orientation. Therefore, the outflow occurrence rate reduces to $f_\textrm{outflow} = f_\textrm{broad}$.

One additional interesting point to consider is how outflow detection rates depend on both SNR and spectral resolution. \cite{XuY25} found that when degrading the high-resolution NIRSpec spectra to match medium-resolution spectra, their outflow detection rate decreased from 30$\%$ to 10$\%$ and they had a bias for faster outflows. In addition, \cite{Xu25} shows that a higher SNR $>$ 50 is needed to robustly detect outflows from emission lines. Therefore, given our high--SNR cut ($>$ 50 on [\oiii] and \ha) and higher spectral resolution ($R$ $\sim$ 3300--3620), our detection rate of outflows ($\sim$ 30\%) is likely more complete than previous studies at Cosmic Noon. For example, \cite{Weldon24} reports an outflow incidence rate of only 10$\%$ at Cosmic Noon with a much lower SNR cut (average SNR $\sim$ 25). As higher-redshift studies typically use lower SNR cuts to maximize sample sizes, outflow measurements tend to have larger uncertainties. This is reflected in the increased scatter in scaling relations (e.g., in Figures \ref{fig:vmax scaling}--\ref{fig: eta vs mass}). 

\begin{figure}
    \centering
    \includegraphics[width=1\linewidth]{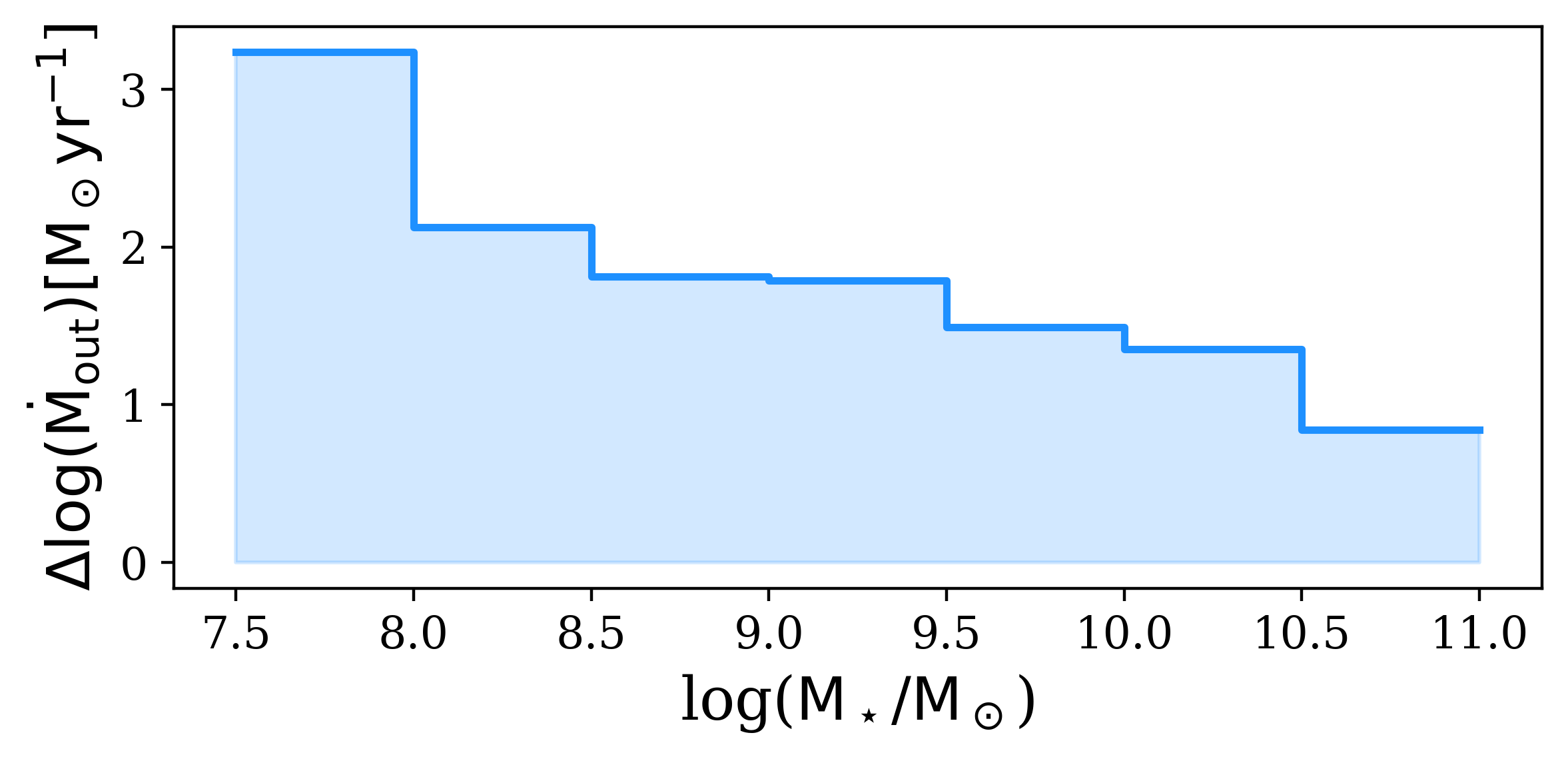}
    \caption{Offset in mass outflow rate at Cosmic Noon from local redshift as a function of log(\Mstar). Lower mass galaxies at Cosmic Noon have much stronger outflow rates than local galaxies in the same mass range.}
    \label{fig:mdot offset}
\end{figure}

\subsection{Redshift evolution of outflows}
\label{sec:redshift evo}
One of the main questions this study aims to address is whether outflow scaling relations evolve over redshift. Here, we further discuss the trends we see in \Vmax, \MdotOut, and $\eta_m$. 

In Figure \ref{fig:vmax scaling}, we found that outflows at $z>2$ exhibit higher velocities than at $z<1$. The maximum outflow velocity for our sample spans around 200--800 \kms, similar to the findings of higher-redshift studies that use \ha\ and [\oiii] as tracers of warm-ionized gas \citep[e.g.,][]{Carniani24, XuY25, Cooper25}. With these higher velocities, we also observe increased mass outflow rates in Figure \ref{fig:mdot vs mass}. We notice that galaxies above $z \gtrsim 2$ have higher \MdotOut\ at lower masses when compared to local redshifts. To evaluate this point, we calculate the average offset between \MdotOut\ for $z=2-3$ samples from the $z<1$ best-fit line for multiple stellar mass bins. Figure \ref{fig:mdot offset} shows a larger offset at lower masses. To ensure this offset is statistically significant, we implement a 2D KS test for Figure \ref{fig:mdot vs mass} between the low and intermediate redshift samples. We use \texttt{2DKS}, a tool by \cite{2DKStest} based on the algorithm first suggested by \cite{Peacock1983}. The KS p-value between these two samples is $9.91 \times 10^{-15}$, which indicates that the low-$z$ sample comes from a different population than the intermediate-redshift sample.

The large offset at lower masses implies that young, lower-mass galaxies at Cosmic Noon host stronger outflows than galaxies with the same stellar mass today. This is consistent with the fact that Cosmic Noon is the peak for cosmic star formation. We notice that this offset decreases with increasing stellar mass and can be attributed to the fact that local massive (log($M_\star/M_\odot$) $>$ 10)  galaxies are more evolved, and may have existed
at Cosmic Noon. Another explanation for this decrease is that the outflowing gas in more massive galaxies is likely to be more suppressed by gravitational and hot CGM confinement.

Despite a higher amount of gas expelled, there is little to no redshift evolution for the warm-phase (T$\approx 10^4$ K) mass loading factor ($\eta_m$). The efficiency and amount of gas carried out by the outflow are regulated by the SFR. At the same time, SFR is positively correlated with stellar mass and, by proxy, gravitational potential. At all redshifts, the warm-phase outflows dominate at lower masses, but drop off at higher masses due to gravitational and hot CGM pressure confinement. This suggests that the feedback effects between star formation and gravitational potential remain fairly consistent throughout cosmic time. However, it is important to note that we are measuring current star formation ($\sim10$ Myr, inferred from \ha) rather than the epoch of star formation that drove the observed outflows. 

\section{Conclusion} 
\label{sec:conclusion}
In this paper, we identify 98 new galaxies with outflow signatures from a high--SNR galaxy sample in the Keck Baryonic Structure Survey (KBSS) and the Keck Lyman Continuum Spectroscopic Survey (KLCS) at Cosmic Noon. 

\begin{itemize}
    \item The overall detection rate of outflows is 30$\%$ (98 of 323 galaxies with SNR $>$ 50 detections of \ha\ or [\oiii]) based on the existence of the broad component in nebular emission lines (Figure \ref{fig:SFMR} and Section \ref{sec: gaussian}).

    \item To investigate the scaling relations across cosmic time, we construct a joint sample of 387 galaxies from $z \sim$ 0--9, including galaxies with known outflows in the literature and 98 new galaxies from this study. 

    \item On average, our outflow sample hosts higher velocity outflows (\Vmax\ $\approx$  200$-$800 \kms) than local literature samples (\Vmax\ $\approx$ 100$-$500 \kms) for a fixed stellar mass (\Mstar). We find that correlations between \Vmax\ and \Mstar\ are statistically significant, but are less apparent for SFR and \SigmaSFR\ (Figure \ref{fig:vmax scaling}). The larger scatter in the latter can be attributed to uncertainties in the outflow geometry and a smaller dynamic range (Section \ref{sec:outflow velocity}).

    \item The mass loading factor ($\eta_m$=\MdotOut/SFR) of our outflow sample lies between 0.3 and 10. We find an anti-correlation between $\eta_m$ and stellar mass and a positive correlation with \SigmaSFR\ (Figures \ref{fig: eta vs mass} and \ref{fig:eta vs sigmaSFR}). For the latter, it is important to note that our observations reflect the most recent SFR rather than the SFR that likely originated the outflows millions of years prior (Section \ref{sec:massloading}).

    \item The 98$\%$ percentile of outflowing gas is able to travel beyond R$_\textrm{vir}$. However, the majority of outflowing gas traveling below $V_{75}$ does not reach R$_\textrm{vir}$ and has a higher likelihood of being recycled into the central galaxy (Figure \ref{fig:Vfrac vs Mdyn} and Section \ref{sec:bound unbound}).

    \item Outflow measurements are affected by line-of-sight projection effects. Taking a simple disk-like approximation, the mean inclination of our galaxies is $\approx  67.8^\circ$, meaning that the velocities measured are likely not the ``down-the-barrel" velocities. However, it is important to consider that the majority of galaxies in this study may not be disk-like, but rather triaxial, complicating inclination estimates (Section \ref{sec: detectability}).
    
    \item If outflows are bi-conical, then our outflow detection rate of 30$\%$ corresponds to an average outflow opening angle of $\theta > 42^\circ$ and a total outflow occurrence rate of 30$\%$. However, previous KBSS studies have not found significant evidence of such geometry. If galaxies at Cosmic Noon host irregular or spherical outflows, then the outflow occurrence rate would be $f_\textrm{outflow} = f_\textrm{broad} = 30\%$ (Section \ref{sec: detectability}).

    \item For the joint sample, we find a redshift evolution in \MdotOut\ as a function of \Mstar, where lower-mass galaxies at high redshifts expel more mass for a fixed \Mstar, but become consistent with what is observed for local galaxies at higher \Mstar\ (Figure \ref{fig:mdot offset}). This is consistent with the suppression of feedback in massive galaxies by gravitational and hot CGM confinement, regardless of redshift. Another explanation is that local massive galaxies are more evolved and have Cosmic Noon progenitors (Section \ref{sec:redshift evo}).

\end{itemize}

Overall, large samples of high--SNR individual galaxies are critical for making robust measurements of outflow properties. Here, we have presented the largest outflow sample of star-forming galaxies at Cosmic Noon to date using high--SNR spectra from KBSS and KLCS. We find that warm-ionized outflows at Cosmic Noon play a significant role in mass ejection from the galaxy, driven by a heightened mass-outflow rate. The majority of the outflowing gas in a single feedback cycle is likely to remain bound. However, over multiple feedback cycles, a significant fraction of baryons can be lost from their host halos. For future outflow studies beyond $z > 2$, obtaining high--SNR and/or spatially resolved spectra (e.g., from James Webb Space Telescope/NIRSpec IFU) will benefit our understanding of outflow geometry and the properties of the \hii\ regions where they are launched from. 

\begin{acknowledgments}
The data presented herein were obtained at Keck Observatory, which is a private 501(c)3 non-profit organization operated as a scientific partnership among the California Institute of Technology, the University of California, and the National Aeronautics and Space Administration. The Observatory was made possible by the generous financial support of the W. M. Keck Foundation. 

T.L and X.X would like to acknowledge Ryan Cooper, Alberto Saldana-Lopez, and Yi Xu for generously providing their data measurements. In addition, T.L would like to thank John Chisholm for his insightful comments on line-fitting and Adam Miller for his feedback on this study. X.X and T.B.M acknowledges the fellowship funding from the Center for Interdisciplinary Exploration and Research in Astrophysics (CIERA) at Northwestern University. A.L.S is supported by the David and Lucile Packard Foundation (Packard Fellowship, grant 2024-77399). This material is based upon work supported by the National Science Foundation Graduate Research Fellowship Program under Grant No. 2025381248 (N.K.C). C.A.F.G was supported by NSF through grants AST-2108230 and AST-2307327; by NASA through grants 80NSSC22k0809, 80NSSC22K1124 and 80NSSC24K1224; by STScI through grant JWST-AR-03252.001-A; and by BSF through grant $\#$2024262.

Finally, the authors wish to recognize and acknowledge the very significant cultural role and reverence that the summit of Maunakea has always had within the Native Hawaiian community. We are most fortunate to have the opportunity to conduct observations from this mountain.

\end{acknowledgments}

\software{
astropy \citep{2013A&A...558A..33A,2018AJ....156..123A,2022ApJ...935..167A}, 
Matplotlib \citep{Hunter:2007},
NumPy \citep{2020NumPy-Array}, 
pandas \citep{reback2020pandas}
 }

\appendix

\section{Data Tables}
Here, we include tables of galaxy and outflow properties measured from our new Keck/MOSFIRE data (98 galaxies). Please find the tables in the ApJ version of the paper.

\section{Figures}
\begin{figure*}[h]
\label{fig:sfr vs snr}
    \centering
    \includegraphics[width=0.8\linewidth]{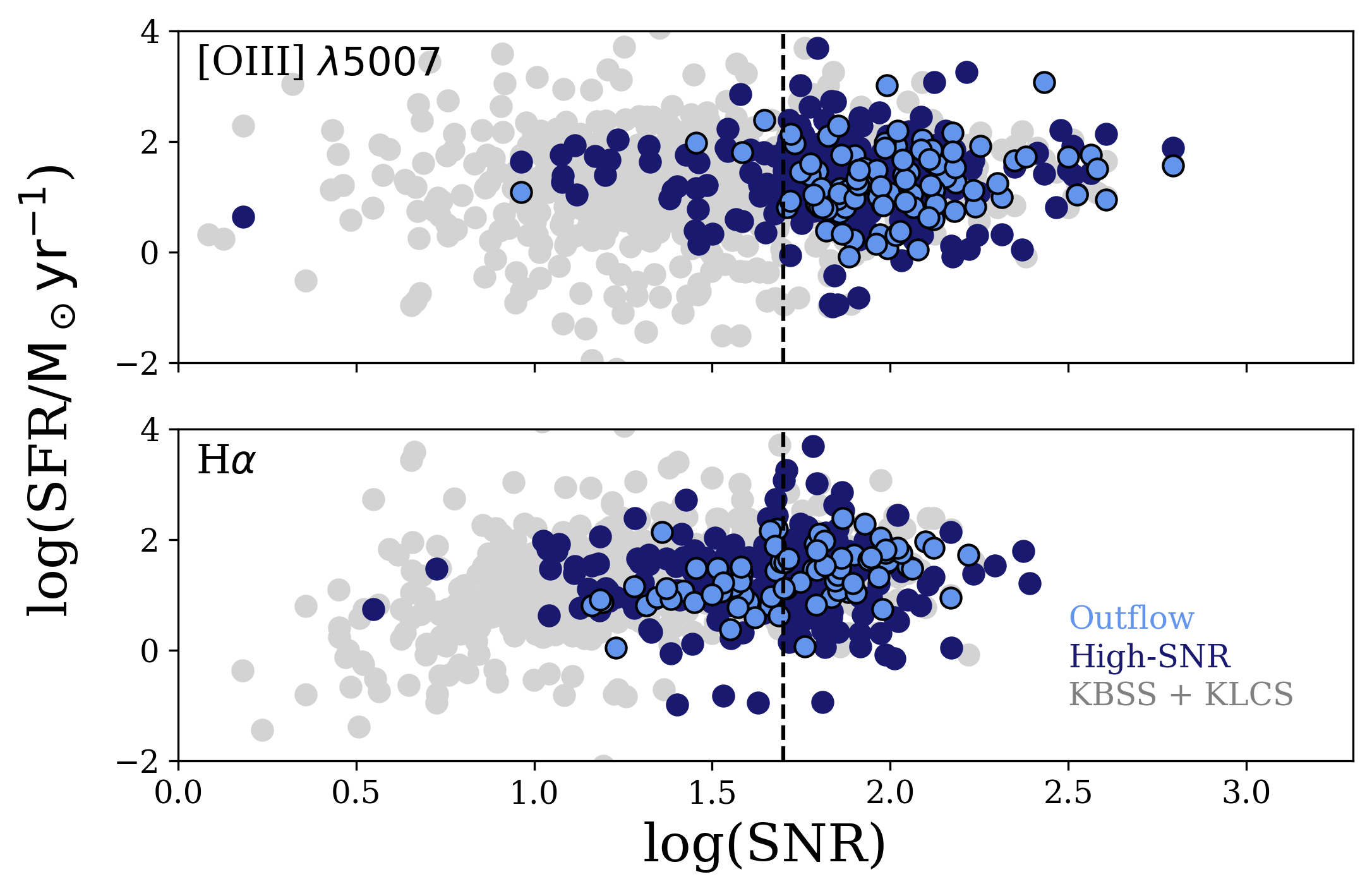}
    \caption{SFR with respect to emission line SNR for [\oiii] $\lambda 5007$ (top) and \ha\ (bottom). In both figures, the light gray dots represent the KBSS+KLCS parent sample, the dark blue dots represent the high--SNR sample (SNR([\oiii])$>$50 or SNR(\ha)$>$50), and the light blue dots represent the final outflow sample. The black dashed line represents our threshold at SNR=50. In both the \ha\ and [\oiii] panels, galaxies left of the dashed line have SNR$>$50 measurements for the other line. Although our outflow sample contains fewer galaxies with SFR$<$1 \Msun yr$^{-1}$, they do not have a strong bias towards galaxies with high SNR observations.}
    
\end{figure*}

\clearpage
\bibliography{literature}{}
\bibliographystyle{aasjournal}

\end{document}